\documentclass{aa}

\usepackage{graphicx}
\usepackage{txfonts}
\usepackage{hyperref}
\usepackage[switch, modulo]{lineno}
\begin{document} 

   \title{Diverse stages of star formation in the IRAS 18162-2048 region}

   \subtitle{Emergence of UV feedback}

   \author{R. Fedriani\inst{1}
        \and G. Anglada\inst{1}
        \and A. Caratti o Garatti\inst{2}
        \and J.F. G\'omez\inst{1}
        \and J. Masqu\'e\inst{3,4}
        \and M. Osorio\inst{1}
        \and B. Stecklum\inst{5}
        \and A.R. Rodr\'iguez-Kamenetzky\inst{6}
        \and R. Galv\'an-Madrid\inst{7}
        \and C. Carrasco-Gonz\'alez\inst{7}
        \and G. Bl\'azquez-Calero\inst{1}
        \and A.F. Placinta-Mitrea\inst{1}
        \and A. Sanna\inst{8}
        \and R. Cesaroni\inst{9}
        \and L. Moscadelli\inst{9}
        \and T.P. Ray\inst{10}
        \and D. Coffey\inst{11}
        \and G.A. Fuller\inst{12,13}
          }

      \institute{Instituto de Astrof\'isica de Andaluc\'ia, CSIC, Glorieta de la Astronom\'ia s/n, E-18008 Granada, Spain\\
              \email{fedriani@iaa.es}
            \and
            INAF, Osservatorio Astronomico di Capodimonte, Salita Moiariello 16, 80131 Napoli, Italy
            \and Departament de F\'isica Qu\`antica i Astrof\'isica (FQA), Universitat de Barcelona (UB), c/ Mart\'i i Franqu\`es 1, 08028 Barcelona, Spain
            \and
            Institut de Ci\`encies del Cosmos (ICCUB), Universitat de Barcelona (UB), c/ Mart\'i i Franqu\`es 1, 08028 Barcelona, Spain
            \and
            Th\"uringer Landessternwarte Tautenburg, Sternwarte 5, 07778 Tautenburg, Germany
            \and
            Instituto de Astronom\'ia Te\'orica y Experimental (IATE), Universidad Nacional de C\'ordoba (UNC), C\'ordoba, Argentina
            \and
            Instituto de Radioastronom\'ia y Astrof\'isica, Universidad Nacional Autónoma de M\'exico, Morelia, Michoac\'an 58089, M\'exico
            \and
            INAF, Osservatorio Astronomico di Cagliari, via della Scienza 5, 09047, Selargius, Italy
            \and
            INAF, Osservatorio Astrofisico di Arcetri, Largo E. Fermi 5, 50125 Firenze, Italy 
            \and
            Dublin Institute for Advanced Studies, School of Cosmic Physics, 31 Fitzwilliam Place, Dublin 2, Ireland
            \and
            University College Dublin, School of Physics, Belfield, Dublin 4, Ireland
            \and
            Jodrell Bank Centre for Astrophysics, Department of Physics \& Astronomy, The University of Manchester, Oxford Road, Manchester M13 9PL, UK
            \and
            I. Physikalisches Institut, Universit\"at zu K\"oln, Z\"ulpicher Str 77, D-50937 K\"oln, Germany
            }

   \date{Received ---; accepted ---}

 
  \abstract
   {High-mass star formation remains a major open problem in astrophysics, particularly regarding the transition between deeply embedded protostars and the onset of ionising radiation capable of producing photodissociation regions (PDRs) and compact \ion{H}{ii} regions.}
   {We aim to characterise the excitation and ionisation conditions of the high-mass star-forming region IRAS~18162$-$2048, which is where the parsec-scale jet HH80$-$81 lies.}
   {We obtained adaptive optics–assisted integral field spectroscopy in the near-IR $K$ band ($1.93-2.47~\mathrm{\mu m}$) with VLT/SINFONI, complemented by VLA X and C bands (3$-$6~cm) and ALMA band~3 ($\sim$3.3~mm) observations. We analysed the continuum and line emission to derive visual extinction and excitation conditions and the kinematics of the gas of the region.}
   {The near-IR continuum reveals two IR sources, IRS~2 and IRS~7, while the main protostellar core, IRAS~18162-2048, remains undetected up to $2.47~\mathrm{\mu m}$. IRS~7 shows a peculiar hydrogen recombination line (HRL) Br$\gamma$ profile with a narrow emission component superimposed on a broad absorption feature, consistent with a B2/B3 zero-age main-sequence (ZAMS) star. Extended H$_2$ emission exhibits a `sawtooth' pattern in the excitation diagram, characteristic of UV radiation in a PDR rather than shock excitation. The radiative transfer model Cloudy reproduces the H$_2$ ro-vibrational populations for $T_\mathrm{gas}=600$~K and $n_\mathrm{H}=7.9\times10^3~\mathrm{cm^{-3}}$. The VLA X and C bands observations reveal a compact radio source previously reported as a `stationary condensation' (SC) and coincident with IRS~7. For the first time, we detect IRS~7/SC in millimetre wavelengths. The spectral index in the 3$-$6~cm and 3.3~mm regime is consistent with optically thin free–free emission.}
    {Our near-IR and radio observations reveal that IRS~7/SC is a B2/B3 ZAMS star that has begun to photo-ionise its environment, giving rise to an extended PDR and a compact \ion{H}{ii} region. The coexistence of this source with the deeply embedded protostar IRAS~18162-2048 and other bubble-like structures in the field suggests a multi-generational star-forming environment. Future \textit{James Webb} Space Telescope observations targeting the H$_2$ pure rotational lines ($3-28~\mathrm{\mu m}$) and other HRLs less affected by extinction will be essential to characterising the cooler molecular and ionised gas to fully disclose the formation history of the region.}


   \maketitle
%

\section{Introduction}\label{seect:intro}
    The formation of low-mass ($M_*<2~M_\odot$) and high-mass ($M_*>8~M_\odot$; $L_\mathrm{bol}>10^3~L_\odot$) stars has traditionally been studied along separate pathways \citep[e.g.][]{beuther2007,frank2014,tan2014}. Yet it is increasingly clear that stars of different masses can form together, evolve on different timescales, and interact dynamically with their common environment \citep[see][for a review]{beuther2025}. In such clustered settings, massive (proto-)stars strongly influence their surroundings through a variety of feedback processes, from powerful jets, winds, and outflows to intense UV stellar radiation that rapidly produces compact \ion{H}{II} regions. Unlike their low-mass counterparts, massive protostars reach the zero-age main sequence (ZAMS) while still accreting \citep{palla1993}, irradiating, and shaping their natal cloud, even during their earliest growth phases. At the same time, several characteristics of star formation appear to scale smoothly across the mass spectrum, such as the ubiquity of accretion discs \citep{beltran2016,andrews2018} and associated collimated jets and outflows \citep{bally2016,anglada2018,ray2021}.

    Feedback from massive stars can disrupt local gas reservoirs, but it also has the potential to induce or accelerate subsequent star formation. Theoretical work and observations over the past decades have pointed towards feedback-driven triggering in bright-rimmed clouds, expanding \ion{H}{II} regions, and wind-blown bubbles \citep[e.g.][]{klein1980,sugitani1989,elmegreen2011}. In addition, protostellar jets and molecular outflows themselves may compress and destabilize dense gas, seeding the next generation of young stars \citep[e.g.][]{girart2001,osorio2017}. Understanding the interplay between accretion, feedback, and clustered formation, especially in high-mass protostellar regions, therefore remains central to building a complete theory of star formation.

    The high-mass star-forming region IRAS~18162-2048 ($d\sim1.4$~kpc, $L_\mathrm{bol}\sim10^4~L_\odot$; \citealp{anez2020}) offers a unique opportunity to investigate these processes in action. Its central main protostar, IRAS~18162-2048, is a deeply embedded $20-30~M_\odot$ protostar driving the spectacular $\sim15$ pc (in projection) HH~80–81 multi-wavelength jet \citep{marti1993,carrasco-gonzalez2010,masque2012,bally2023,mohan2023obs}. The surrounding cluster contains numerous young stellar objects (YSOs) with circumstellar masses (most likely tracing discs) between $0.003$ and $\sim5~M_\odot$, identified in high-resolution ALMA band~6 observations \citep{busquet2019,anez2020}. Mid-IR imaging ($8$–$13~\mu$m) has also revealed embedded sources such as IRS~2 and IRS~7, which may be more evolved sources \citep{stecklum1997}. The region therefore presents a hint of multi-generational star formation, where massive proto-stellar accretion, disc-jet activity, and emerging UV-driven photo-ionisation may coexist within a compact environment.

    Despite the wealth of existing observations, the region IRAS~18162-2048 continues to pose fundamental questions. The origin of the bright H$_2$ emission, the nature of compact centimetre to millimetre sources in the vicinity of the main protostar, and the degree to which newly formed massive stars irradiate and reshape their environment remain unclear. Further, whether these signatures are solely the result of shock activity from ongoing accretion and outflows or they instead mark the ignition of young OB-type stars and the development of compact \ion{H}{II} regions and photodissociation regions (PDRs) is also unknown. To address these questions, we combined near-IR integral field spectroscopy with radio observations to map the molecular and ionised gas with the aim of identifying the emission mechanisms. 
    
    The paper is structured as follows: In Sect.~\ref{sect:obs} we detail the observations taken and the data reduction procedures. In Sect.~\ref{sect:results} we present our results, while in Sect.~\ref{sect:discussion} we discuss the implications of our findings. Section~\ref{sect:conclusions} contains our conclusions.

\section{Observations}\label{sect:obs}

    \subsection{VLT/SINFONI}
    
    The Very Large Telescope (VLT) Spectrograph for INtegral Field Observations in the Near-Infrared \citep[SINFONI;][]{sinfoni2003} data were obtained on 3 August 2018 as part of the programme ID 0101.C-0317(A) (PI: R. Fedriani), targeting the first $\sim10000$~au of the high-mass star-forming region IRAS~18162-2048. The observations consist of eight AB nodding positions, each with DIT=30~s and NDIT=6 for the science positions and NDIT=3 for the sky positions. Total exposure times of 180~s and 90~s were set per frame for the science and sky positions, respectively. The field of view (FoV) was centred at the IRS~7 source covering $8\arcsec\times8\arcsec$ ($11200\times11200$~au$^2$ at a distance of 1.4~kpc) with a spaxel size of 250~mas pixel$^{-1}$, and the position angle was east of north of zero degrees. The observations were adaptive optics assisted with a natural guide star, reaching an angular resolution of $\sim0.6\arcsec$. The natural guide star used was 2MASS J18191071-2047423 with R = 13.6~mag and B-R colour 0.17, and it was separated from the science target by 20\arcsec. We set up the K grating with a spectral resolution of $\sim4000$ ($\sim75~\mathrm{km~s^{-1}}$).
    The data were reduced using the European Southern Observatory pipeline \texttt{gasgano} with standard parameters. A wavelength accuracy of $0.09~\AA$ (i.e. $\sim1.3~\mathrm{km~s^{-1}}$) was achieved using arc lamps (neon plus argon in the K band). The standard star Hip~088201 was used for spectral calibration and telluric correction.

    \subsection{VLA}

    The radio centimetre continuum emission from the IRAS~18162-2048 region was observed with the \textit{Karl G. Jansky }Very Large Array (VLA) of the National Radio Astronomy Observatory (NRAO) in A configuration (VLA-A) at C (covering the frequency range 4–8~GHz, i.e. $\sim$6~cm) and X (covering 8–12~GHz, i.e. $\sim$3~cm) bands during August 2019 (project code: 19A-321, PI: A.R. Rodr\'iguez-Kamenetzky). The phase centre was set at R.A. (J2000) = $18^\mathrm{h}19^\mathrm{m}12.09^\mathrm{s}$, Dec. (J2000) = $+20^\circ{}47\arcmin30.89\arcsec$.

    Data calibration was conducted with the Common Astronomy Software Applications \citep[\texttt{CASA}; version 6.4.1;][]{casa2022} package using the NRAO pipeline optimised for VLA continuum observations. We adopted the standard frequency setup for the continuum, utilising 2 MHz-wide channels across the full bandwidth of each band. The source 3C 286 served as both the bandpass and flux calibrator, while complex gain calibration (phase and amplitude) was derived from frequent observations of J1911–2006. Calibrated datasets from both bands were concatenated into a single file for subsequent imaging and self-calibration. Imaging was carried out with the \texttt{tclean} task in \texttt{CASA}, employing multi-frequency synthesis (nterms = 2) and multi-scale cleaning algorithms \citep{rau2011}. Primary beam correction was applied, and super-uniform weighting was used to maximise angular resolution, yielding a restoring beam of $0.30\arcsec\times0.14\arcsec$ with a position angle of $21.5^\circ{}$ and a root mean square (rms) noise $\sigma=7~\mu$Jy beam$^{-1}$.

    \subsection{ALMA}

    Observations were carried out with the Atacama Large Millimeter/submillimeter Array (ALMA) in band 3 on 27 June 2014 as part of the program 2012.1.00441.S (PI: R. Galván-Madrid). The source J1733-1304 was used as a flux-scale and bandpass calibrator, and J1832-2039 was used as a gain calibrator. Visibility calibration and imaging were performed with \texttt{CASA} version 5.0.0, using as the phase centre R.A. (J2000) = $18^\mathrm{h}19^\mathrm{m}12.094^\mathrm{s}$, Dec. (J2000) = $-20^\circ{}47\arcmin30.91\arcsec$. The spectral setup included four spectral windows of 0.94 GHz of bandwidth each, centred at 92.03, 92.87, 105.89, and 106.73 GHz. All the line-free channels of these spectral windows were combined to produce the continuum image presented in this paper.
    
    The image has a central frequency of 99.39 GHz ($\sim$3.3~mm), a beamsize of $0.99\arcsec\times0.95\arcsec$, a position angle of 77.4$^\circ{}$ (Briggs weighting with robust=0.5), and an rms noise $\sigma=18~\mu$Jy beam$^{-1}$, which was achieved after a few iterations of phase self-calibration.

\section{Results}\label{sect:results}

\subsection{Spatial and spectral distribution of the emission in the near-infrared}

\begin{figure*}[!htb]
        \centering
        \includegraphics[width=0.24\textwidth]{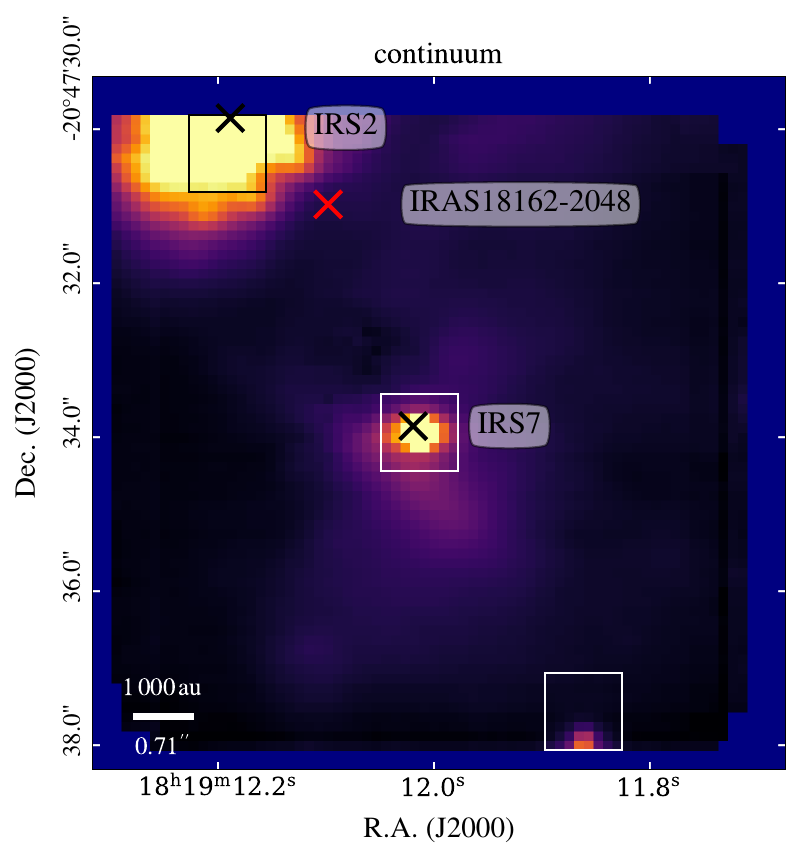}
        \includegraphics[width=0.24\textwidth]{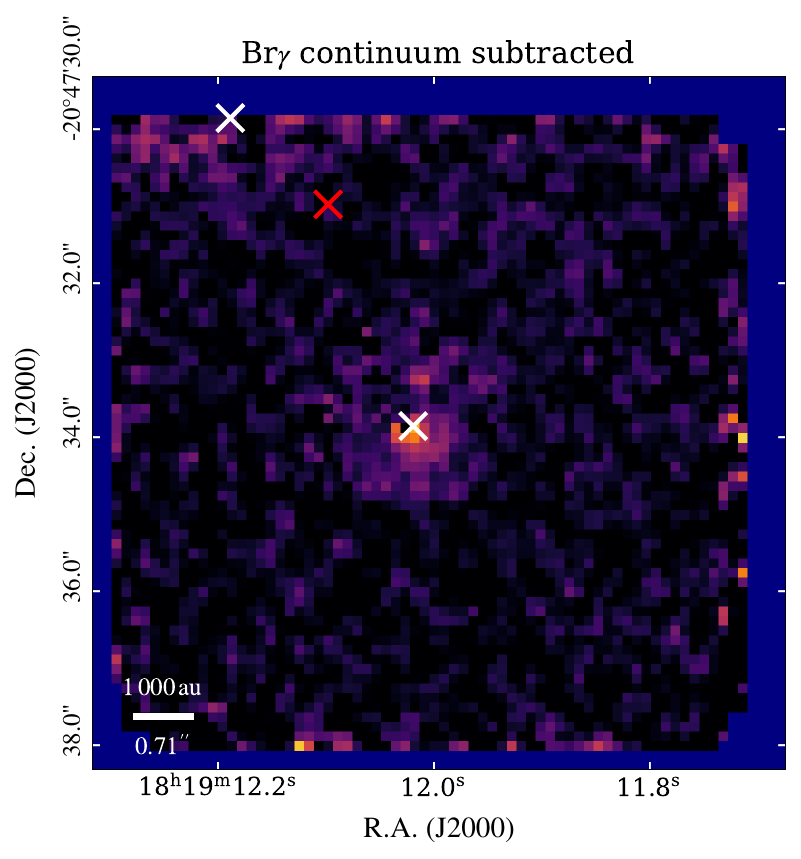}
        \includegraphics[width=0.24\textwidth]{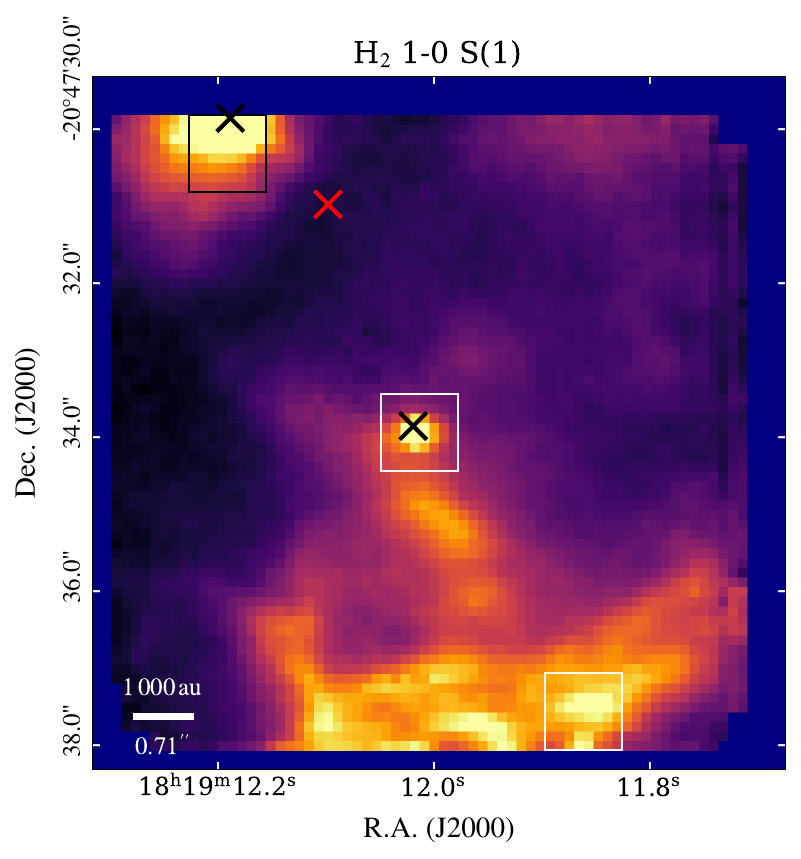}
        \includegraphics[width=0.24\textwidth]{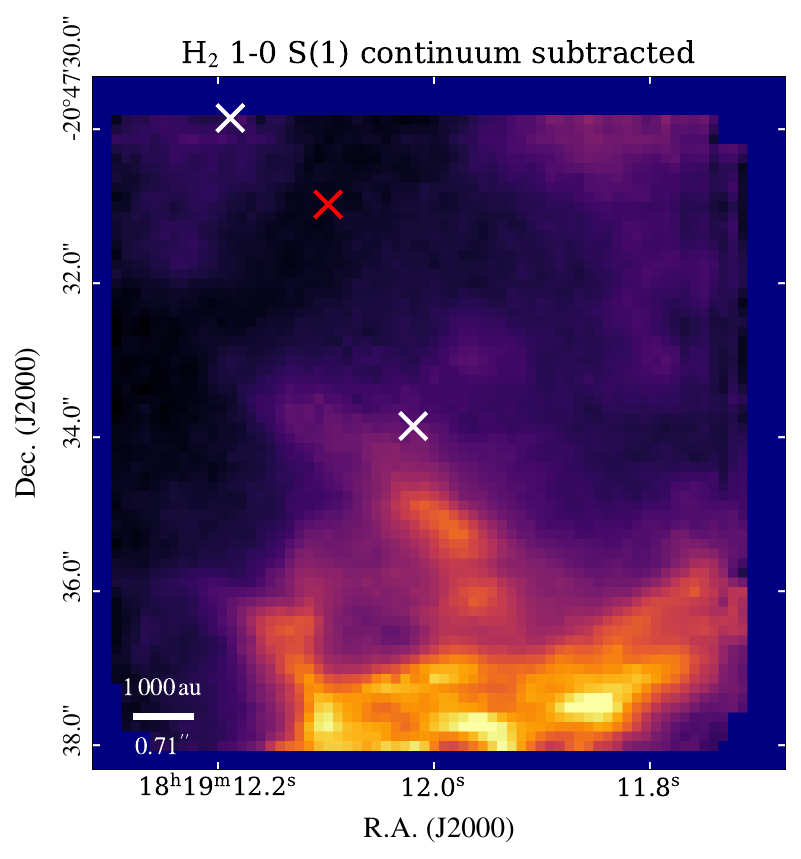}
        
        \caption{\textit{Left:} SINFONI $K$ band continuum view of the IRAS~18162-2048 star-forming region. The black and white boxes represent the extraction regions for the spectra shown in Fig.~\ref{fig:spectra}. The black crosses represent the positions of IRS~2 and IRS~7 (peak intensity pixel from SINFONI), and the red cross shows the position of IRAS~18162-2048 (from the VLA data). \textit{Middle left:} Peak intensity continuum-subtracted map for the Br$\gamma$ emission line. The white crosses represent the positions of IRS~2 and IRS~7, and the red cross indicates the position of IRAS~18162-2048. \textit{Middle right:} Peak intensity map for the $1-0$~S(1) H$_2$ emission line including the continuum emission. \textit{Right:} Same as middle-right panel but with the continuum subtracted. In all panels, the full FoV of SINFONI is shown.}
        \label{fig:cont_H2_BrG}
    \end{figure*}

    Our spectro-imaging observations cover the first $\sim$10000~au of the IRAS~18162-2048 region. The left panel of Fig.~\ref{fig:cont_H2_BrG} shows the continuum emission constructed using 555 line-free channels across the SINFONI $K$ band spanning $1.93-2.47~\mathrm{\mu m}$. We identified two previously reported sources, IRS~2 and IRS~7 \citep{yamashita1987,tamura1991}. The main protostar IRAS~18162-2048 is not detected at any wavelength within this spectral range, even when summing all continuum channels, indicating a much higher level of extinction toward this source. We observed extended continuum emission around IRS~7, suggesting scattered light in the vicinity. We measured its $K_\mathrm{s}$ band magnitude by integrating the flux at $2.159~\mathrm{\mu m}$. We first adopted the $2.8\arcsec$ aperture defined by \citet{stecklum1997}, obtaining $K_\mathrm{s}=11.52\pm0.01$~mag, consistent with the value reported at the epoch of March 1995 ($\sim11.60$~mag; see their Table 1). Moreover, 2MASS reports $K_\mathrm{s}\sim11.51$~mag at the epoch of 10 June 1999 \citep{skrutskie2006}. This suggests no major brightening or dimming events have occurred over the past few decades, based on the consistent brightness level across all three measurements, although the intervening activity remains unknown. Using our improved angular resolution, we measured the magnitude in a smaller region of $1\arcsec$ (central white box in Fig.~\ref{fig:cont_H2_BrG}), yielding $K_\mathrm{s}=12.79\pm0.02$~mag. No magnitude was estimated for IRS~2, as it lies near the edge of our FoV, and it is clearly not entirely covered in our observations, leading to flux losses.
        
\begin{figure*}[!htb]
        \centering
        \includegraphics[width=0.9\textwidth]{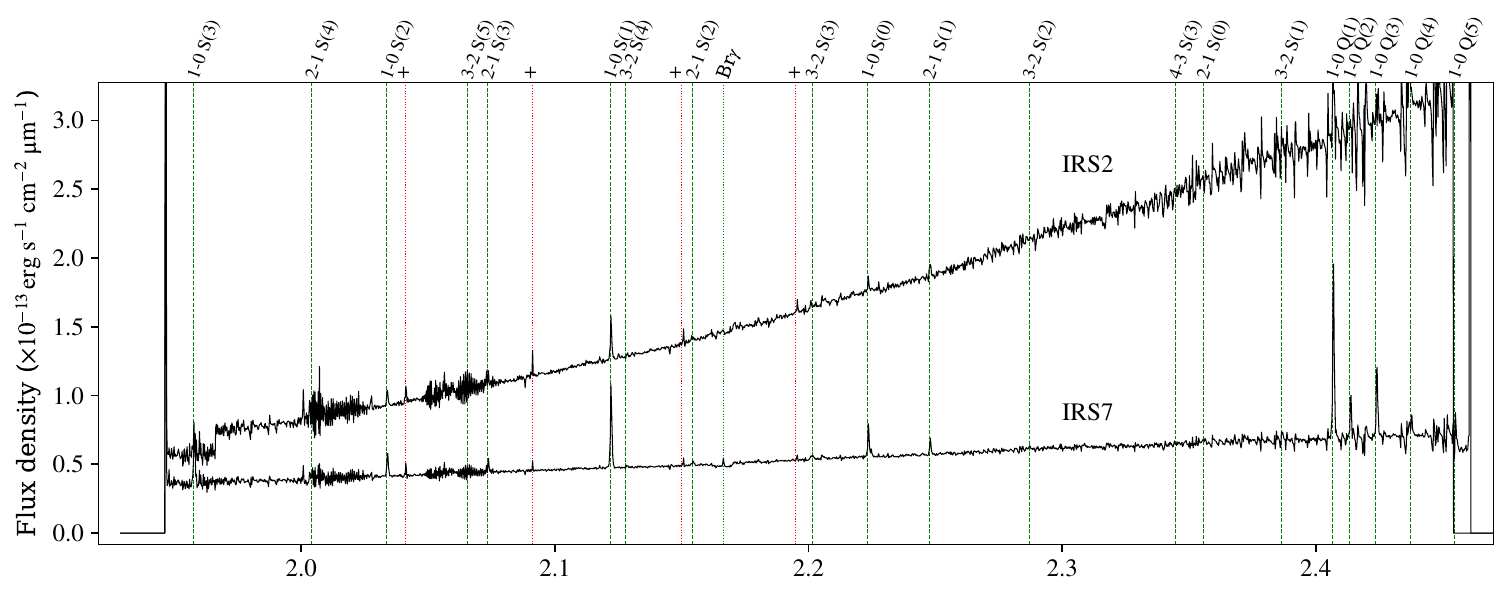}
        \includegraphics[width=0.9\textwidth]{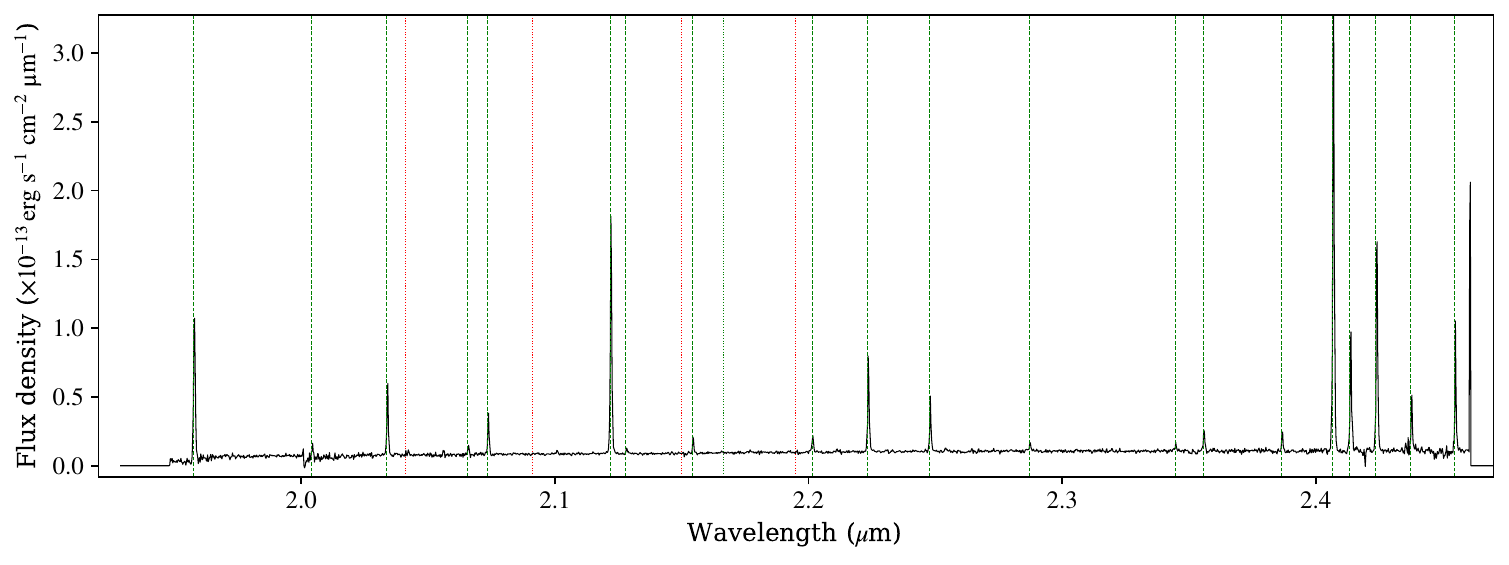}
        
        \caption{\textit{Top:} Spectra for IRS~2 and IRS~7 as labelled. \textit{Bottom:} Spectrum extracted at the highest S/N in the main region of H$_2$ emission. Extraction boxes are shown in Fig.~\ref{fig:cont_H2_BrG}. Main H$_2$ emission lines are marked as vertical dashed green lines, Br$\gamma$ as a dotted green line, and noisy peaks as dotted red lines and labelled as `$+$'.}
        \label{fig:spectra}
    \end{figure*}

    The middle-right panel of Fig.~\ref{fig:cont_H2_BrG} shows the combined line and continuum emission of the $1-0$~S(1) molecular hydrogen (H$_2$) transition at $2.12~\mathrm{\mu m}$. The continuum-subtracted H$_2$ emission (Fig.~\ref{fig:cont_H2_BrG}, right panel) resembles what is typically associated with protostellar outflows \citep[see e.g.][]{varricatt2010,caratti2015}, even displaying a bow-shaped-like feature \citep[see e.g.][]{fedriani2018,ray2023,crowe2024,crowe2025}. However, as discussed in Sect.~\ref{sect:H2_PDR}, this emission is produced by UV pumping and most likely traces a PDR with a bubble-like structure. Conversely, the hydrogen recombination line (HRL) Br$\gamma$ at $2.16~\mathrm{\mu m}$ is detected exclusively towards the position of IRS~7 (see Fig.~\ref{fig:cont_H2_BrG} middle-left panel and Fig.~\ref{fig:spectra} top panel). In star-forming regions, the Br$\gamma$ line is often used as a tracer of accretion and ejection processes occurring very close to the central protostar \citep{coffey2010,alcala2014,caratti2016}. In our case, however, the emission is not spatially extended, or it presents a P-Cygni profile and therefore is unlikely to trace protostellar jets \citep{davies2010,fedriani2019} or the innermost region of accretion discs given its line profile \citep{garcia-lopez2020,garcia-lopez2024}, although this depends on the inclination angle. A detailed analysis of the Br$\gamma$ line profile and its possible origin is presented in Sect.~\ref{sect:BrG}.

    Figure~\ref{fig:spectra} shows the spectra extracted from the regions marked by the black and white boxes in Fig.~\ref{fig:cont_H2_BrG}. Both IRS~2 and IRS~7 display clear signs of reddening, but it is stronger in IRS~2 (see top panel of Fig.~\ref{fig:spectra}). The IRS~2 spectrum shows no prominent emission lines other than the ubiquitous $1-0$~S(1) H$_2$ line, which is detected across the entire FoV. \citet{aspin1994CO} also found a featureless spectrum for IRS~2. In contrast, the IRS~7 spectrum exhibits numerous H$_2$ transitions, most likely arising from a PDR. There are a number of spurious peaks in the spectra marked with dotted red lines and labelled with `$+$'. It should also be noted that the wiggles in the range 2.0 to 2.1 $\mu$m are noise and not real emission. Notably, the HRL Br$\gamma$ displays a composite absorption plus emission profile. A close-up view of this feature is provided in Fig.~\ref{fig:BrG_zoom}. The bottom panel of Fig.~\ref{fig:spectra} presents the rich H$_2$ spectrum extracted from the white south box region shown in Fig.~\ref{fig:cont_H2_BrG}. Gaussian profiles were fitted to the H$_2$ emission lines, and the fluxes were derived by integrating under the fitted curves. The rms noise was estimated from the continuum surrounding each line to determine the associated flux uncertainties. Table~\ref{tab:observed_lines_IRAS18} summarises the H$_2$ lines detected within the white south box region, which represents the highest signal-to-noise ratio (S/N) region. For each transition, the table lists the vacuum wavelength, the integrated flux, and the corresponding S/N. Interestingly, nowhere in our cube did we find evidence of additional emission lines that are usually associated with accretion and ejection processes in the circumstellar environment of YSOs, such as [\ion{Fe}{ii}], \ion{Na}{i}, \ion{He}{i}, or the CO band head.

    \begin{table}
    \caption{Molecular hydrogen (H$_2$) emission lines detected on IRAS~18162-2048. }
    \label{tab:observed_lines_IRAS18}      
    \centering          
    \begin{tabular}{c c c c}    
    \hline\hline       
    \noalign{\smallskip}
    Transition & $\lambda_\mathrm{vac}$ & Flux & S/N \\ 
    &($\mu\mathrm{m}$) &  ($\times10^{-15}\mathrm{~erg~cm^{-2}~s^{-1}}$) & \\
    \noalign{\smallskip}
    \hline              
    \noalign{\smallskip}
      $1-0$~S(3) & 1.95755 & $5.08\pm0.06$ & 85.3 \\
      $2-1$~S(4) & 2.00407 & $0.23\pm0.02$ & 11.8 \\
      $1-0$~S(2) & 2.03375 & $1.80\pm0.04$ & 41.2 \\
      $3-2$~S(5) & 2.06556 & $0.31\pm0.04$ & 7.4 \\
      $2-1$~S(3) & 2.07351 & $0.87\pm0.03$ & 32.4 \\
      $1-0$~S(1) & 2.12182 & $6.89\pm0.02$ & 288.1 \\
      $3-2$~S(4) & 2.12796 & $0.11\pm0.02$ & 4.5 \\
      $2-1$~S(2) & 2.15422 & $0.32\pm0.02$ & 13.6 \\
      $3-2$~S(3) & 2.20139 & $0.46\pm0.04$ & 12.6 \\
      $1-0$~S(0) & 2.22330 & $2.63\pm0.03$ & 103.6 \\
      $2-1$~S(1) & 2.24772 & $1.14\pm0.03$ & 45.3 \\
      $3-2$~S(2) & 2.28702 & $0.24\pm0.03$ & 7.7 \\
      $4-3$~S(3) & 2.34448 & $0.16\pm0.03$ & 6.5 \\
      $2-1$~S(0) & 2.35563 & $0.58\pm0.04$ & 13.1 \\
      $3-2$~S(1) & 2.38645 & $0.58\pm0.05$ & 11.6 \\
      $1-0$~Q(1) & 2.40659 & $14.5\pm0.05$ & 299.5 \\
      $1-0$~Q(2) & 2.41343 & $3.89\pm0.05$ & 83.0 \\
      $1-0$~Q(3) & 2.42372 & $6.88\pm0.05$ & 149.3 \\
      $1-0$~Q(4) & 2.43749 & $1.61\pm0.05$ & 35.5 \\
      $1-0$~Q(5) & 2.45475 & $2.91\pm0.04$ & 79.3 \\
    \noalign{\smallskip}
    \hline                  
    \end{tabular}
    \tablefoot{Fluxes were measured within the region indicated by the white south box in Fig.~\ref{fig:cont_H2_BrG}.}
    \end{table}

\subsection{The hydrogen recombination line Brackett gamma}\label{sect:BrG}

    \begin{figure*}[!htb]
        \centering
        \includegraphics[width=0.95\textwidth]{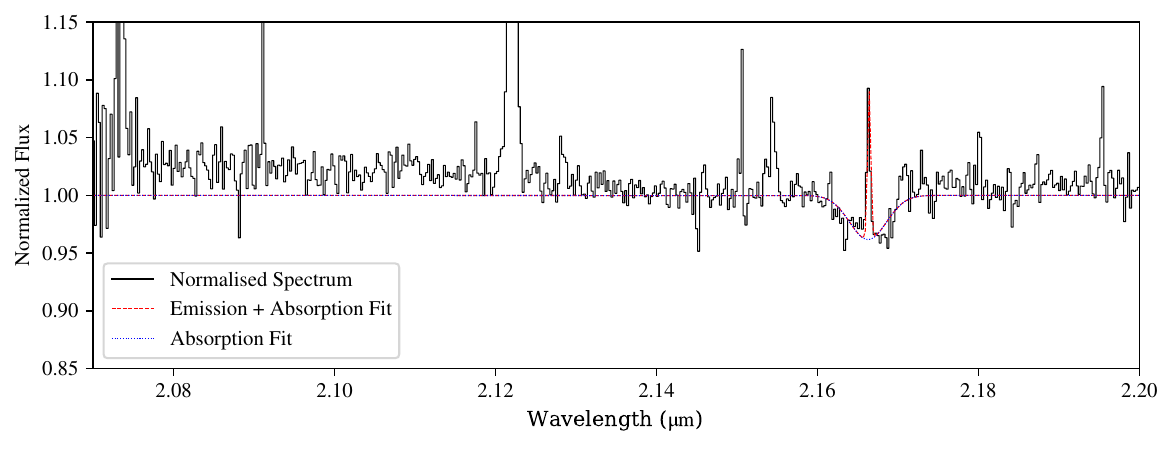}
        
        \caption{Zoom-in of the IRS~7 spectrum at the $2.07-2.20~\mathrm{\mu m}$ range. A double Gaussian profile (emission plus absorption) fit to the Br$\gamma$ line is shown as a dashed red line, whereas the dotted blue line shows the single Gaussian profile fit to the absorption component. Strong emission lines correspond to H$_2$, in particular $2-1$~S(3), $1-0$~S(1), $3-2$~S(4), and $2-1$~S(2), at 2.074, 2.122, 2.128, and 2.154~$\mathrm{\mu m}$, respectively.}
        \label{fig:BrG_zoom}
    \end{figure*}
    
    Line profiles can provide key diagnostics of the physical mechanisms at work in YSOs. For example, inverse P-Cygni profiles (i.e. strong red-shifted continuum absorption) are associated with infalling material in molecular emission at large scales \citep{zapata2008}, whereas P-Cygni profiles (i.e. strong blue-shifted continuum absorption) are linked to strong ejection activity for ionised winds close to the central sources \citep{najarro1997}. The Br$\gamma$ line in YSOs has been observed to trace jets close to the protostar \citep{caratti2016} and accretion processes within circumstellar discs \citep{garcia-lopez2020}. A prominent P-Cygni profile was observed in the high-mass YSO IRAS~13481$-$6124 \citep{fedriani2018}, where the line traces powerful jets close to the massive protostar. In our case, we detected Br$\gamma$ emission only at the position of IRS~7, showing a peculiar line profile. The line exhibits a clear emission core superimposed on a broad absorption feature (see Fig.~\ref{fig:BrG_zoom}). We fitted a double Gaussian profile to the Br$\gamma$ line, accounting for both the emission and absorption components. The emission component is unresolved at our spectral resolution of $\sim75~\mathrm{km~s^{-1}}$ and peaks at $\sim0~\mathrm{km~s^{-1}}$ in the local standard of rest. These two facts suggest that it could be tracing an \ion{H}{ii} region, while the absorption feature displays a full width at half maximum of $\sim750~\mathrm{km~s^{-1}}$. Considering only the absorption component, we estimate an equivalent width (EW) of $2.2\pm0.2~\AA$. To the best of our knowledge, this is the first detection of an HRL in the region of IRAS~18162-2048.

    The observed line profile is characteristic of an OB-type ZAMS star. \citet{bik2005b} conducted a high spectral resolution ($\mathcal{R}\sim10000$, corresponding to $\sim30~\mathrm{km~s^{-1}}$) survey of deeply embedded young massive star candidates and established that O- and B-type stars can be empirically distinguished by their Br$\gamma$ EW, with a boundary at $EW=5~\AA$, and by the presence (or a lack) of a number of emission lines. Although our EW measurement could nominally place IRS~7 in the O-type regime by the classification scheme of \citet{bik2005b}, this value may be affected by veiling, which tends to reduce the apparent depth of absorption features. We can confidently rule out an O-type classification for IRS~7, as we did not detect the characteristic \ion{C}{iv} lines at 2.069, 2.078, and 2.083~$\mathrm{\mu m}$; the \ion{He}{i} lines at 2.058, 2.112, and 2.113~$\mathrm{\mu m}$; the \ion{N}{iii} line at 2.115~$\mathrm{\mu m}$; nor the \ion{He}{ii} line at 2.185~$\mathrm{\mu m}$, all of which were found in the spectra of O-type sources by \citet{bik2005b}. We therefore classify IRS~7 as a B2/B3 ZAMS star \citep[see][and references therein for details on the classification scheme]{bik2005b}. This was confirmed later by our radio observations (see Sect. \ref{sect:radio}).

    We interpret the broad absorption component as originating from the stellar photosphere \citep[][found full width at zero intensity of $800~\mathrm{km~s^{-1}}$ for the Br$\gamma$ for other sources in their survey]{bik2005b}, while the narrow unresolved emission core most likely traces ionised gas associated with an incipient \ion{H}{ii} region surrounding IRS~7. This interpretation is consistent with the presence of extended continuum and H$_2$ emission in the vicinity, which together indicate that IRS~7 is already influencing its immediate environment through both radiative and mechanical feedback.

\subsection{Extinction and excitation conditions}\label{sect:H2_PDR}

    Line ratios involving transitions that originate from the same upper energy level can be used to estimate the extinction along the line of sight \citep[e.g.][]{nisini2008}. Ideally, these lines should be widely separated in wavelength to maximise sensitivity to reddening. However, our spectral setup covers only the $K$ band, from 1.93 to 2.47~$\mathrm{\mu m}$. To estimate the visual extinction, $A_V$, we used the H$_2$ $1-0$~Q(3)/$1-0$~S(1) line ratio, which offers the highest S/N among the available line pairs. We followed the same approach described by \citet{costa2022} (but see also \citet{zannese2025}) and adopted the near-IR extinction law from \citet{rieke1985}.

\begin{figure*}[!htb]
        \centering
        \includegraphics[width=0.33\textwidth]{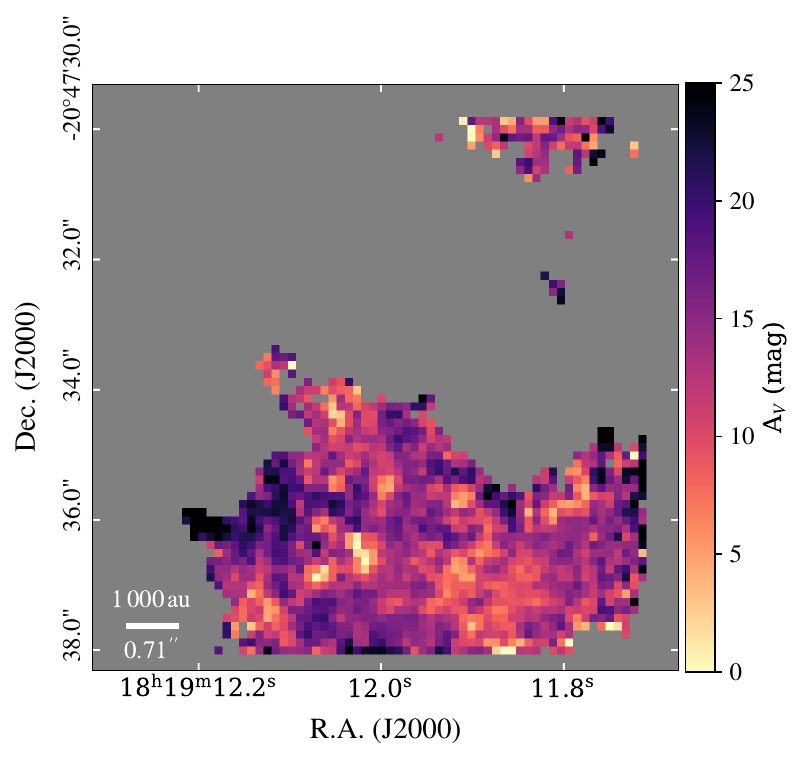}
        \includegraphics[width=0.33\textwidth]{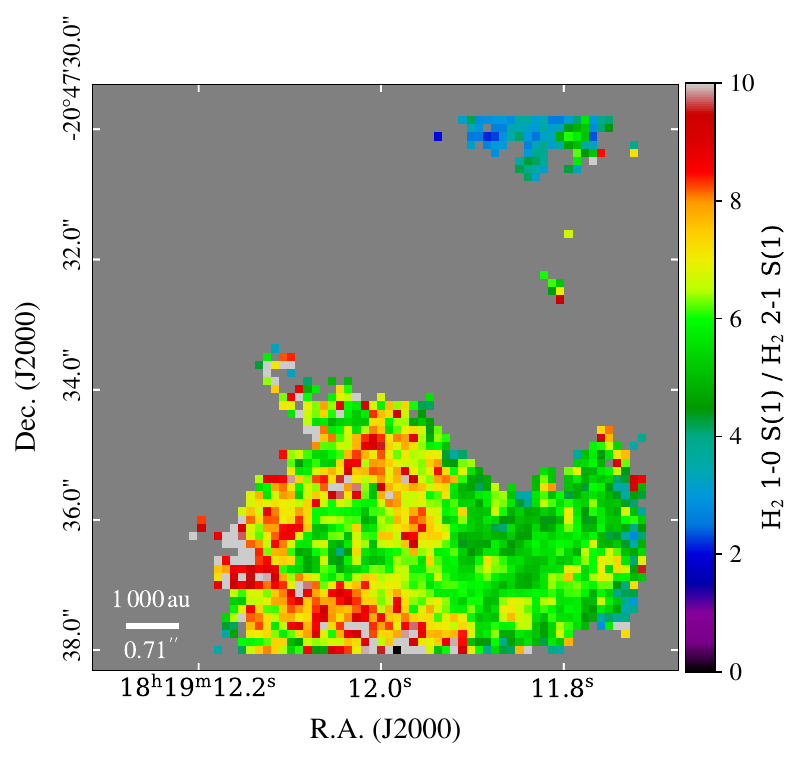}
        \includegraphics[width=0.33\textwidth]{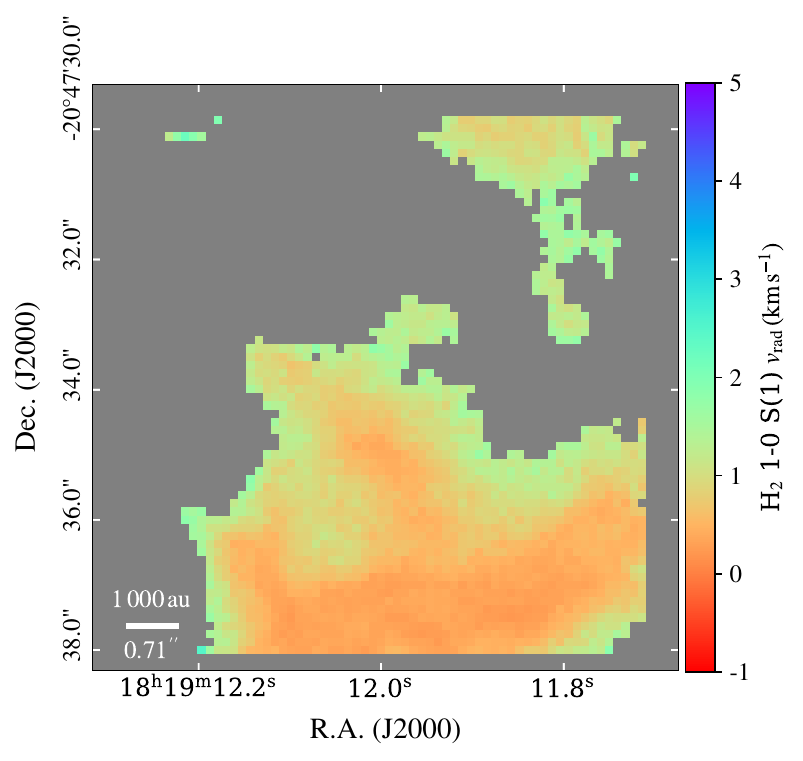}

        \caption{\textit{Left:} Extinction map obtained from the H$_2$ $1-0$~Q(3)/$1-0$~S(1) line ratio. \textit{Middle:} Excitation map from the H$_2$ line ratio $1-0$~S(1)/$2-1$~S(1). \textit{Right:} Radial velocity map using the H$_2$ $1-0$~S(1) line. In all panels, grey indicates masked pixels with S/N$<3$.}
        \label{fig:extinction_excitation}
    \end{figure*}

    Since our data provide spectra for every pixel, we could construct a spatially resolved $A_V$ map. Figure~\ref{fig:extinction_excitation} (left panel) shows the resulting pixel-by-pixel extinction distribution. For instance, at the position of IRS~2, the $1-0$~Q(3) line is not detected with sufficient S/N, leading to unreliable extinction estimates in that region, and therefore it has been masked (grey region in the figure). Across the main H$_2$ bow-shaped structure, the average extinction is $\sim12$~mag, consistent with the value of $12.4$~mag reported by \citet{stecklum1997} using an independent method. In their method, the authors reddened the blackbody spectrum of a B2 ZAMS star by varying the extinction until the model matched the observed photometry from the R to the K band.

    The H$_2$ line ratio $1-0$~S(1)/$2-1$~S(1) provides insights into the excitation mechanism of the emitting gas (Fig.~\ref{fig:extinction_excitation} middle panel). For shock excitation, this ratio is expected to exceed $\sim10$, whereas for purely UV-pumped emission, it is typically $\sim1.7$ \citep{burton1992}. After correcting the fluxes for extinction using our derived $A_V$ values, we obtained an average ratio of $\sim$5.9. This clearly indicates that the H$_2$ emission in the region is inconsistent with purely shock-driven thermal excitation and instead points to UV fluorescence as the dominant process (see Sect.~\ref{sect:rv_diag}). However, \citet{burton1990} showed that in dense PDRs, collisional processes can alter the level populations, leading to deviations from those expected for purely UV-excited gas (see also \citealt{sternberg1999}). Therefore, we are probably observing an intermediate case between UV-excited and shocked H$_2$. This is more evident in the southeast region of our observations, where the ratio is greater than eight, although with just a few pixels with ratios consistent with shocks (i.e. $\gtrsim10$).

    Velocity maps provide key insight into the kinematics associated with a given emitting species. The right panel of Fig.~\ref{fig:extinction_excitation} presents the radial velocity map of the H$_2$ $1-0$~S(1) line with respect to the systemic velocity of the region, which is the strongest and best isolated transition in our dataset. We used the Python package \texttt{bettermoments} \citep{teague2018} to fit Gaussian profiles to the line emission across the field, considering only pixels with S/N$>3$. The average radial velocity within the main emitting region is $\sim1~\mathrm{km~s^{-1}}$, indicating negligible bulk motion along the line of sight. It should be noted that although the nominal spectral resolution of our observations is $\Delta v \sim 75~\mathrm{km,s^{-1}}$, the centroid velocity of the line can be measured with higher precision, scaling as $\Delta v / \sqrt{\mathrm{S/N}}$. For the H$_2$ $1-0$ S(1) line, with an average $\mathrm{S/N}\sim100$, this corresponds to uncertainties of $\sim7.5~\mathrm{km,s^{-1}}$. Even for the pixels with S/N$\sim3$, the uncertainty would be $\sim43~\mathrm{km,s^{-1}}$. In any case, such low velocities are consistent with two scenarios: (i) a potential outflow oriented nearly in the plane of the sky, resulting in minimal radial velocity components or (ii) PDR-excited H$_2$ emission. In shock–driven H$_2$ emission associated with protostellar jets, velocity shifts of tens to hundreds of kilometres per second are typically observed \citep[e.g.][]{fedriani2018,fedriani2019} and so is evidence of blue- and red-shifted components \citep[e.g.][]{fedriani2020}. In contrast, PDR–excited H$_2$ usually exhibits very small velocity gradients close to the systemic velocity \citep{burton1990,marconi1998}. Combined with the excitation analysis presented earlier, this velocity structure strongly supports a PDR origin for the observed H$_2$ emission rather than shock-driven outflow activity (see also Sect.~\ref{sect:rv_diag}). Together, the extinction, excitation, and velocity maps analyses provide a consistent picture in which UV radiation from nearby massive (proto-)stars plays a key role in shaping the observed near-IR emission.

\subsection{H$_2$ ro-vibrational diagrams}\label{sect:rv_diag}

    The combined effects of UV radiation and collisional excitation produce two limiting regimes for the ro-vibrational level populations. In dense and/or hot environments, such as shocked gas in protostellar outflows, frequent collisions drive the level populations towards a thermal (Boltzmann) distribution, resulting in ro-vibrational diagrams that follow a straight line in a logarithmic excitation plot \citep[e.g.][]{caratti2008}. In contrast, low-density gas exposed to intense UV radiation exhibits a distinct distribution, with populations that do not decrease monotonically with excitation energy. In excitation diagrams, this manifests as a characteristic `sawtooth' pattern, which serves as a diagnostic signature of UV-pumped H$_2$ emission in PDRs \citep{burton1992,kaplan2017,le2017}. Such diagrams provide a clear distinction between collisional and radiative excitation mechanisms and give constraints on both the gas temperature and the ortho-to-para ratio (OPR).

\begin{figure}[!htb]
        \centering
        \includegraphics[width=0.49\textwidth]{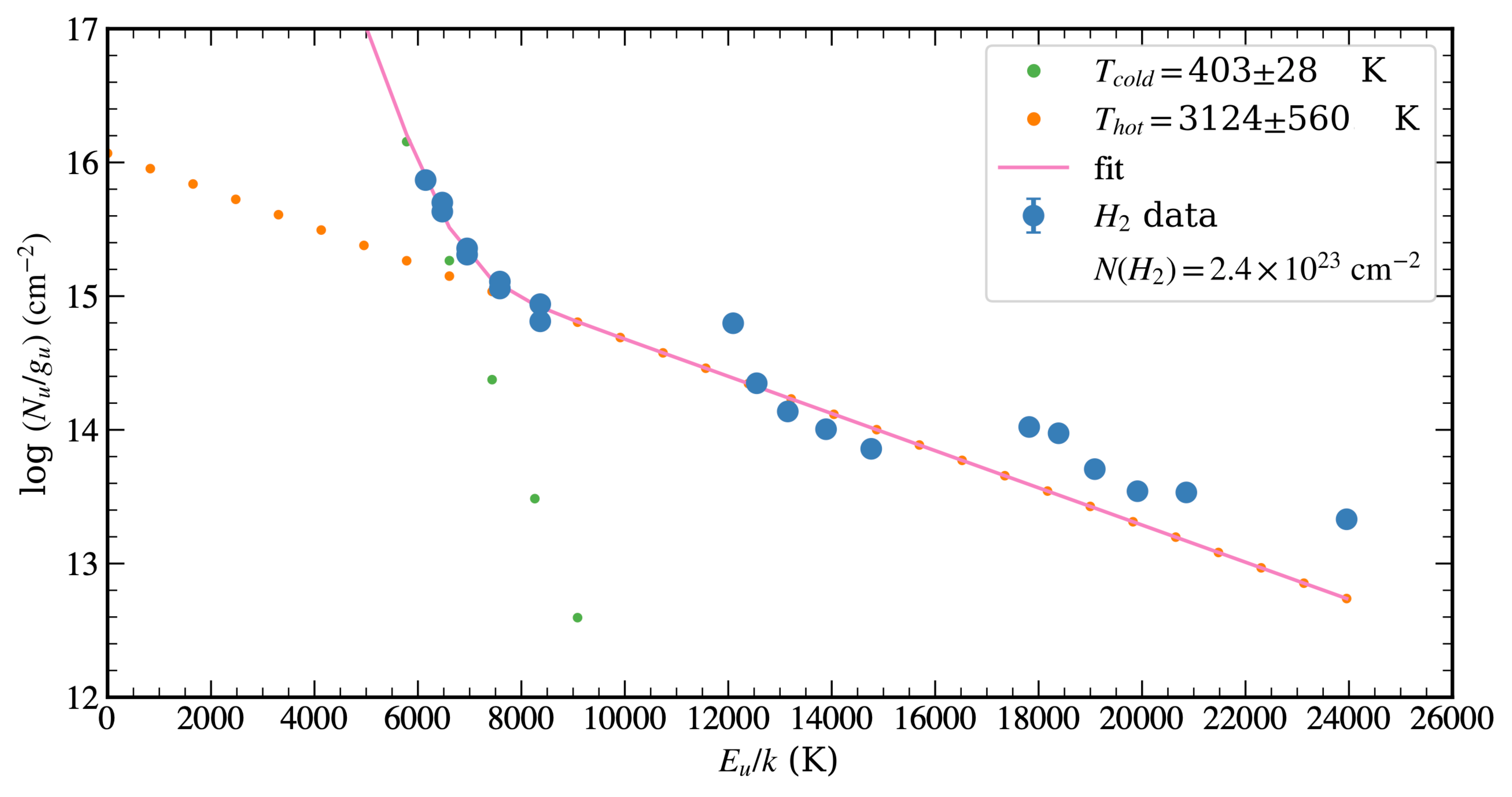}

        \caption{Ro-vibrational diagram using the fluxes extracted from the white south box depicted in Fig.~\ref{fig:cont_H2_BrG} and summarised in Table~\ref{tab:observed_lines_IRAS18}. The blue circles represent the observed H$_2$ data after extinction correction ($A_V=12$~mag), and the pink line shows the two-temperature component. The plot was generated using the PDR Toolbox \citep{pound2023}.}
        \label{fig:rv_diag}
    \end{figure}

    We constructed a ro-vibrational diagram using the de-reddened fluxes from Table~\ref{tab:observed_lines_IRAS18} corresponding to the white south box, adopting $A_V=12$~mag. To model the emission, we fitted a two-temperature component while leaving the OPR as a free parameter. A single temperature component fit failed to reproduce the observed H$_2$ intensities. Figure~\ref{fig:rv_diag} shows the resulting ro-vibrational diagram for the observed H$_2$ lines together with the fit obtained using the PDR Toolbox \citep{pound2023}. The characteristic `sawtooth' pattern of PDRs is clearly visible, confirming the dominance of UV pumping in the emission, instead of a near-linear thermal distribution. The best-fit model yields a cold component with $T_\mathrm{cold}\sim400$~K, a hot component with $T_\mathrm{hot}\sim3000$~K, a total H$_2$ column density $N(\mathrm{H}_2)\sim2\times10^{23}~\mathrm{cm^{-2}}$, and an OPR$=2.76\pm0.14$.

    The derived hot-component temperature is higher than typically observed in protostellar outflow knots, where values of $T_\mathrm{hot}\sim2000$~K are more common, and the ro-vibrational diagrams usually follow a near-linear thermal distribution \citep{caratti2015,costa2022}. This further supports the conclusion that UV pumping is the dominant excitation mechanism in the H$_2$ emission for which $T_\mathrm{cold}$ is a better tracer of the kinetic temperature of the gas. The fitted OPR is $2.76\pm0.14$, which is somewhat high for pure PDR emission (see e.g. \citealt{le2017}, and references therein). As mentioned above, we are probably observing an intermediate case between UV- and shocked-excited H$_2$ emission. However, the collective near-IR evidence points towards a strong contribution from UV-pumped excitation.

    \subsection{Radio emission at millimetre and centimetre wavelengths}\label{sect:radio}

\begin{figure}[!htb]
        \centering
        \includegraphics[width=0.45\textwidth]{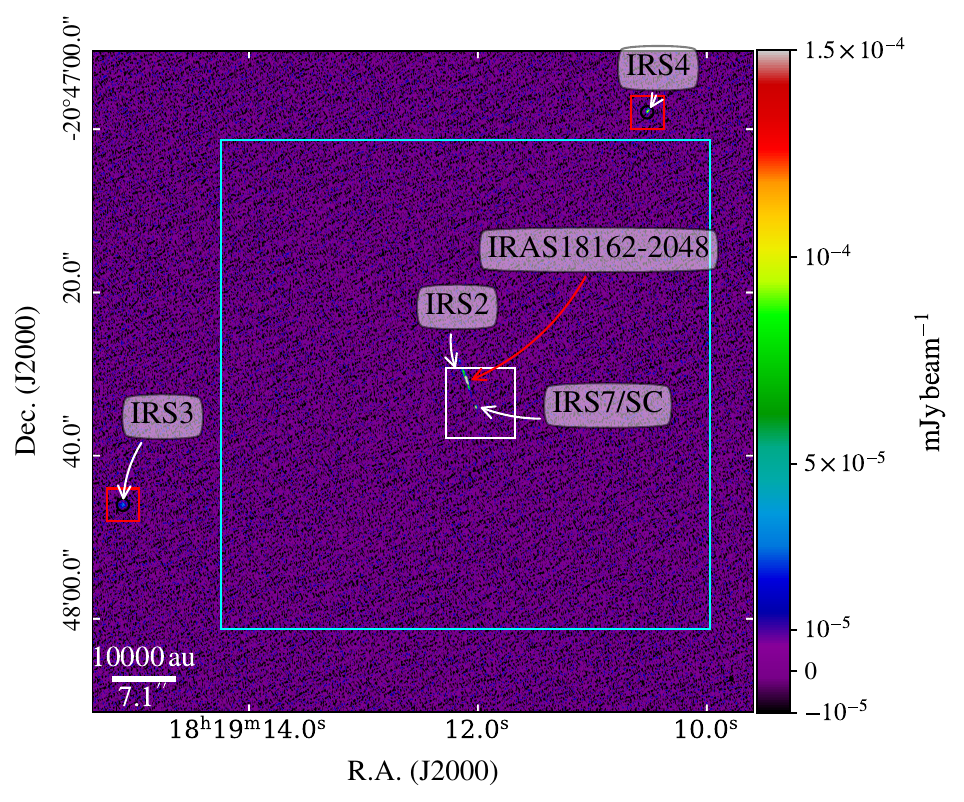}

        \includegraphics[width=0.45\textwidth]{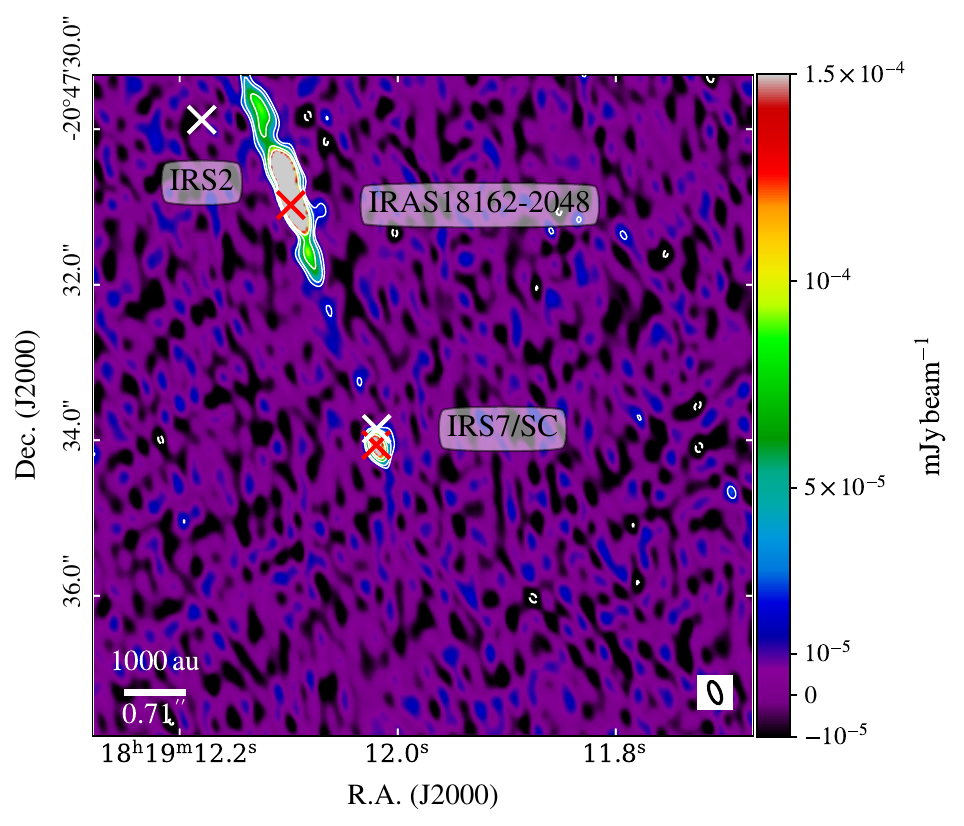}
        \includegraphics[width=0.24\textwidth]{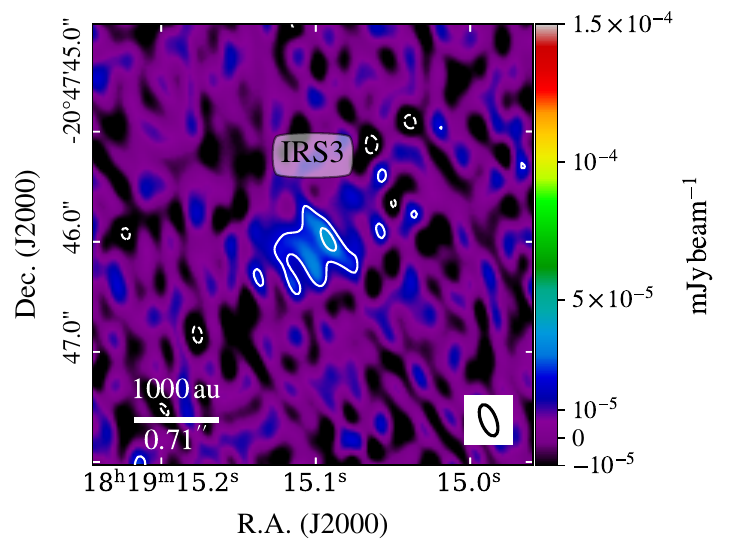}
        \includegraphics[width=0.24\textwidth]{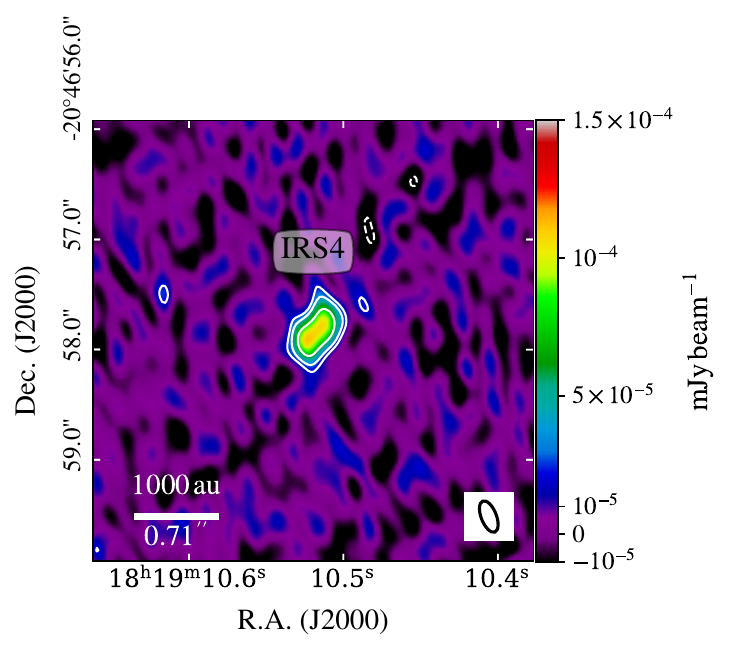}
        \caption{\textit{Top:} VLA X and C band continuum image. The cyan box represents the field shown in Fig.~\ref{fig:alma_band3}, whereas the white box represents the FoV of SINFONI. The main sources in the region are labelled. \textit{Middle:} Same as above but zoomed in on the SINFONI FoV. Contours are $(-3, 3, 5, 10, 15) \times\sigma_{X+C}$ where $\sigma_{X+C}=
        7\mu$Jy~beam$^{-1}$. The red crosses represent the VLA positions for IRAS~18162-2048 and IRS~7/SC, and the white crosses are the SINFONI positions for IRS~2 and IRS~7/S. \textit{Bottom:} Zoomed in views of the red boxes in the top panel highlighting IRS~3 (left) and IRS~4 (right). The synthesised beams are shown at the bottom-right corners.}
        \label{fig:vla_cx}
    \end{figure}

    We present new high-sensitivity VLA A-configuration observations at $\sim$3–6~cm (X and C bands). Figure~\ref{fig:vla_cx} shows the resulting radio continuum map of the IRAS~18162-2048 region, where several compact sources were clearly detected. In addition to the emission associated with the main protostar IRAS~18162-2048, we also detected a non-resolved compact radio counterpart to IRS~7. Further emission was observed at the positions of the IR sources IRS~3 and IRS~4 originally identified by \citet{yamashita1987}. Both objects were later detected in the radio by \citet{marti1993}, who catalogued them as sources~20 (IRS~3) and~12 (IRS~4) in their Table~2.
    
    Interestingly, both IRS~3 and IRS~4 are visible in the H$\alpha$ (6564~\AA) image presented in Fig.~1 of \citet{bally2023}, suggesting the presence of ionised material in relatively low-extinction regions. Moreover, each source appears as a point-like object surrounded by bubble-like structures, corresponding to the molecular hydrogen objects (MHOs) 2357 and 2359, respectively \citep[see Figs.~15 and 16 in][]{bally2023}. A similar identification was made by \citet{mohan2023obs}, who classified IRS~3 and IRS~4 as `YSO2' and `YSO1', connected to their `Reg3' and `Reg5' regions (see their Fig.~2). IRS~4, in particular, has been proposed to be a Herbig Ae/Be star of spectral type B2, and it is thought to be responsible for carving the prominent H$_2$ bubble observed around it \citep{bally2023}. We performed 2D Gaussian fits to all compact detections except for the main protostar IRAS~18162-2048 and its jet, as analysis of its complex morphology goes beyond the scope of this paper. Table~\ref{tab:vla} presents a summary of their positions and flux densities. A discussion of the nature of these sources is provided in Sect.~\ref{sect:discussion}.

    \begin{table}
    \caption{VLA X and C band detections in the region IRAS~18162-2048, excluding the main protostar and its jet. }
    \label{tab:vla}      
    \centering          
    \begin{tabular}{l c c c}    
    \hline\hline       
    \noalign{\smallskip}
    Source & R.A. (J2000) & Dec. (J2000) & $S_\mathrm{3-6~cm}$ \\ 
    &(hh:mm:ss.sss) &  (dd:mm:ss.ss) & (mJy)\\
    \noalign{\smallskip}
    \hline              
    \noalign{\smallskip}
    IRS~4$^{a,\dagger}$ & 18:19:10.518 & -20:46:57.85 & $0.366 \pm 0.031$ \\
    IRS~7$^{b,\dagger,*}$ & 18:19:12.018 & -20:47:34.08 & $0.283 \pm 0.015$ \\
    IRS~3$^{a,\dagger}$ & 18:19:15.102 & -20:47:46.04 & $0.370 \pm 0.065$ \\
    \noalign{\smallskip}
    \hline                  
    \end{tabular}
    \tablefoot{Positional uncertainties are $\sim50$~mas. The symbols are as follows: $^a$ catalogued in Table~1 from IR observations by \citet{yamashita1987}; $^b$ catalogued in Table~5 from IR observations by \citet{tamura1991}; $^{\dagger}$ catalogued in Table~2 as sources 12, 13, and 20, respectively, from radio observations by \citet{marti1993}; $^*$ referred to as stationary condensation (SC) in \citet{marti1995}.}
    \end{table}

\begin{figure}[!htb]
        \centering
        \includegraphics[width=0.45\textwidth]{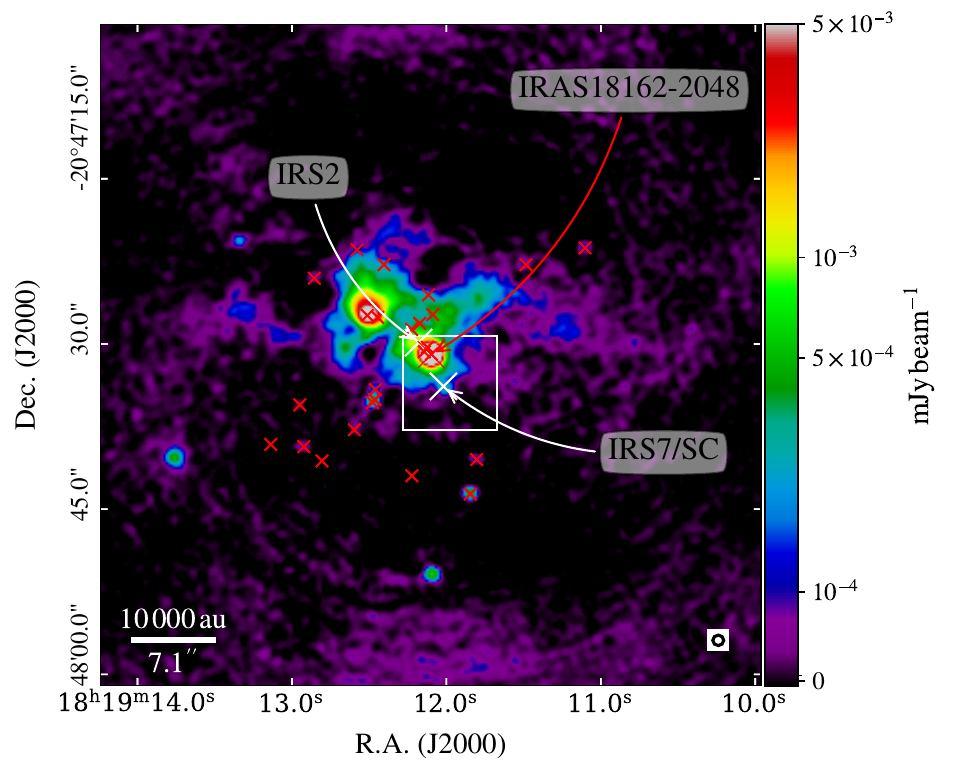}
        \includegraphics[width=0.45\textwidth]{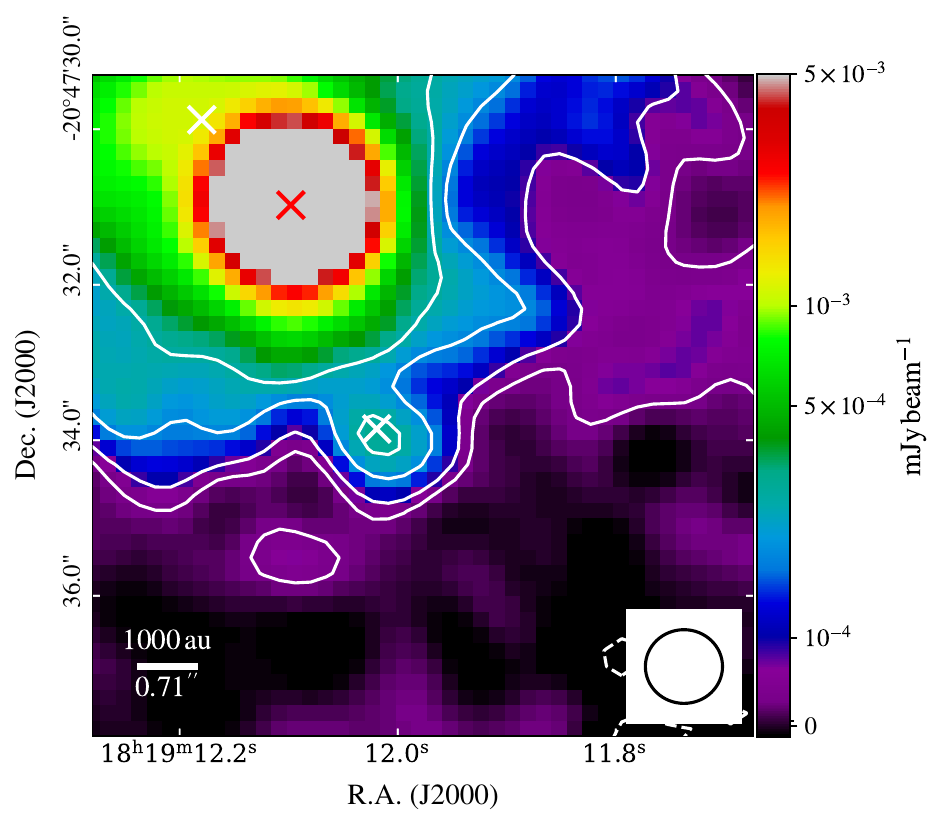}
        \caption{\textit{Top:} ALMA band 3 ($\sim$3.3~mm) image in the region IRAS~18162-2048. Red crosses are ALMA 1.14~mm detections in the field \citep{busquet2019}, with the larger cross corresponding to the main protostar IRAS~18162-2048. The large white crosses indicate the positions of IRS~2 and IRS~7/SC from their SINFONI positions. The white box represents the SINFONI FoV. \textit{Bottom:} Same as above but zoomed in on the SINFONI FoV. Contours are $(-3,3,5,10,15)\times\sigma_\mathrm{3.3mm}$, where $\sigma_\mathrm{3.3mm}=18~\mu$Jy~beam$^{-1}$. The synthesised beams are shown at the bottom right corners.}
        \label{fig:alma_band3}
    \end{figure}

    Figure~\ref{fig:alma_band3} shows complementary ALMA continuum emission at $\sim3.3$~mm. We overlayed the ALMA $\sim1.14$~mm compact sources reported by \citet{busquet2019} as red crosses. Several objects are detected in both bands, while others are only recovered at one of the ALMA frequencies. This behaviour may result from a combination of dust opacity, differing evolutionary stages, and optical depth effects. We also note the considerable difference in angular resolution and sensitivity between the ALMA $\sim1.14$~mm data ($\sim45$~mas, $\sim56~\mu$Jy~beam$^{-1}$; \citealt{busquet2019}) and our $\sim3.3$~mm image ($\sim1\arcsec$, $\sim18~\mu$Jy~beam$^{-1}$), which likely contributes to the observed difference between wavelengths.

    A particularly noteworthy result is the detection of IRS~7 in our 3.3~mm data. It is the first time this source has been detected in the millimetre regime. A Gaussian fit to the ALMA band~3 emission of IRS~7 yields $S_{\rm 3.3mm} = 0.457 \pm 0.310$~mJy, and it is non-resolved. The source was originally identified in the radio continuum by \citet{marti1993}, who later referred to it as the `stationary condensation' (SC) due to its lack of proper motions, in contrast to the extremely high velocity ($\sim1000\,\mathrm{km\,s^{-1}}$) HH~80–81 jet knots \citep{marti1995,bally2023}. IRS~7 was first identified by \citet{tamura1991} in the IR. Later, \citet{stecklum1997} associated IRS~7 with the SC, and our detections confirm this identification.
        
    We calculated the spectral index of IRS~7/SC given their non-resolved detections at both centimetre and millimetre wavelengths, defined as 

    \begin{equation}
        \alpha = \frac{\log_{10}(S_{\nu_1}/S_{\nu_2})}{\log_{10}(\nu_1/\nu_2)}
    ,\end{equation}where $\nu_1=8.19$~GHz and $\nu_2=99.3$~GHz, corresponding to the observed frequencies for VLA and ALMA, respectively. We obtained a value of $\alpha=0.19\pm0.27$, which is consistent with almost optically thin free-free emission (i.e.\ $-0.1\leq\alpha\leq2$ for free-free emission; \citealt{anglada1998,purser2016,sanna2018}). In fact, assuming that the emission at $3-6$~cm arises from free-free emission, extrapolating with a spectral index of $-0.1$ (optically thin emission) would provide a strict lower limit to the free-free contribution at higher frequencies. This extrapolation would yield a value of $0.220\pm0.011$~mJy at 3.3~mm, which is what we actually detect within the errors. Moreover, \citet{marti1995} reported a flat spectral index between 3.6 and 6~cm, consistent with optically thin free-free emission as well. Thus, we can confidently conclude that dust is not the dominant origin of the emission in IRS 7/SC, supporting the interpretation that this source is a compact photo-ionised \ion{H}{ii} region powered by a B2/B3 ZAMS star (see Sect.~\ref{sect:discussion}).

    We note that IRS~2 is not detected at either $3-6$~cm or 3.3~mm. This applies to our data and to previous studies \citep[e.g.][]{marti1993,rodriguez-Kamenetzky2017,busquet2019}. This non–detection may suggest a later evolutionary stage, where accretion and jet activity have declined and no detectable ionised region has developed, or alternatively a lower–luminosity source below our sensitivity threshold. It could also correspond to a non-stellar object, with the observed IR emission arising from scattered light from nearby sources, as suggested by \citet{aspin1994CO}.

\section{Discussion}\label{sect:discussion}

\subsection{IRS~7: A B2/B3 zero-age main-sequence star powering an \ion{H}{ii} and a photodissociation region}\label{sect:discussion_irs7}

    Our results strongly indicate the presence of sources of UV photons within the region of IRAS~18162-2048. The Br$\gamma$ line profile observed towards IRS~7/SC is consistent with that of a B2/B3 ZAMS star \citep{bik2005b}. In addition, the presence of a PDR is revealed by the H$_2$ emission morphology, its kinematics (Fig.~\ref{fig:extinction_excitation}), and the distinct ro-vibrational `sawtooth' pattern in the excitation diagram (Fig.~\ref{fig:rv_diag}). The measured $1-0$~S(1)/$2-1$~S(1) ratio and radial velocity maps (Fig.~\ref{fig:extinction_excitation}) further rule out a shock origin for the molecular emission, confirming that UV-pumped excitation dominates. The collective near-IR evidence thus points to a strong source of UV photons in the vicinity. Interestingly, the main source in the region, IRAS~18162-2048, remains undetected up to $2.47~\mathrm{\mu m}$ (Fig.~\ref{fig:cont_H2_BrG}), likely due to the extremely high extinction towards its core. This suggests that IRS~7/SC is the main contributor to the observed PDR emission, at least in the near-IR regime.

    We note that IRAS~18162-2048 has been extensively observed at millimetre and centimetre wavelengths. The optical component was first identified by \citet{gyulbudaghian1978}, while early radio observations by \citet{rodriguez1980} detected free-free emission in the region. Later, \citet{marti1993} presented high-resolution VLA observations at 6~cm, revealing the SC for the  first time \citet{marti1995}. Unlike the HH~80–81 radio knots, which exhibit proper motions exceeding $1000~\mathrm{km~s^{-1}}$ \citep{marti1995,bally2023}, the SC source has remained stationary for decades, indicating that it is not part of the main protostellar jet.

\begin{figure}[tb]
        \includegraphics[width=0.5\textwidth]{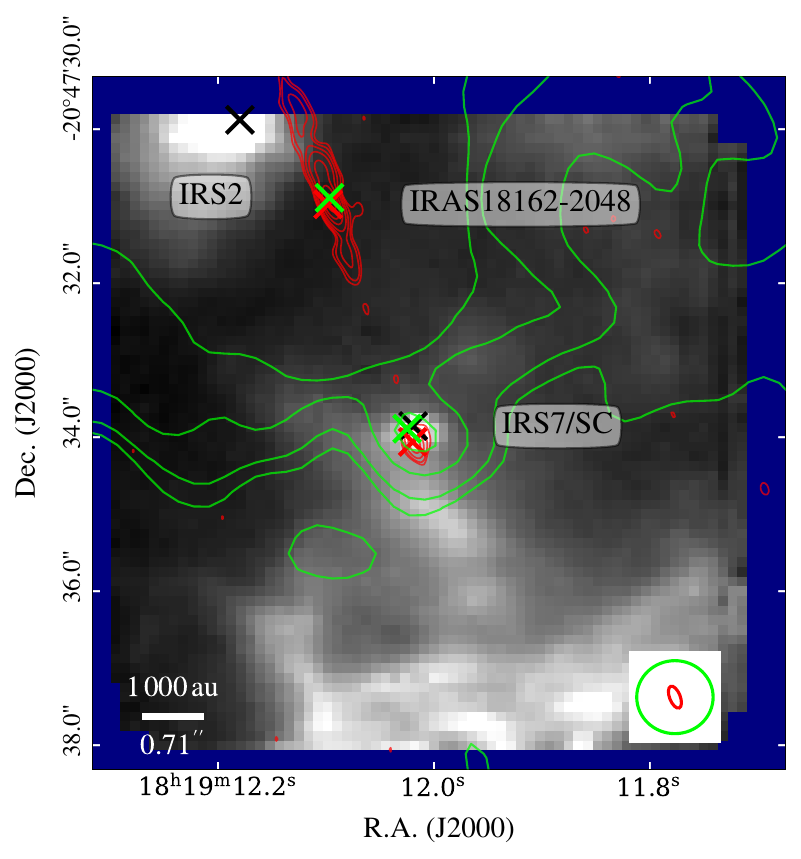}

        \caption{Inner 10000~au view of the IRAS~18162-2048 high-mass star-forming region. The greyscale image is H$_2$ 1-0 S(1) at $2.12~\mathrm{\mu m}$. Red contours are VLA X and C bands in the A configuration. Contours are $(3, 5, 10, 20, 50, 100, 200) \times\sigma_{X+C}$, where $\sigma_{X+C}=7\mu$Jy~beam$^{-1}$. Green contours are ALMA $\sim3.3$~mm.  Contours are $(3, 5, 10, 15) \times\sigma_{3.3\mathrm{mm}}$, where $\sigma_{3.3\mathrm{mm}}=18\mu$Jy~beam$^{-1}$. Black crosses correspond to SINFONI positions. The synthesised beams are shown at the bottom-right corner.}
        \label{fig:nir_vs_radio}
    \end{figure}

    For the first time, IRS~7/SC has been detected in the millimetre regime using our ALMA $\sim3.3$~mm observations. The resulting spectral index, $\alpha = 0.19 \pm 0.27$, is consistent with optically thin free–free emission and supports the presence of a compact \ion{H}{ii} region. Based on the radio continuum alone, it is not possible to conclusively distinguish between shock–ionised jet emission and photo-ionised gas. When the near-IR and radio diagnostics are considered together, photo-ionisation becomes the most plausible interpretation. Specifically, the absence of measurable proper motions, the compact morphology, the measured spectral index, the Br$\gamma$ line profile, and the H$_2$ excitation properties all favour photo-ionised gas rather than shock-excited jet emission for this source. Figure~\ref{fig:nir_vs_radio} shows the H$_2$ $1-0$~S(1) emission overlaid with both the VLA $3-6$~cm as red contours and the ALMA $\sim3.3$~mm continuum as green contours. It is also worth noting the non-detection of IRS~7 at 1.14~mm in \citet{busquet2019}. This may point to the fact that no accretion disc is present at this source, implying an older generation of stars in this region.

\begin{figure}[tb]
        \includegraphics[width=0.5\textwidth]{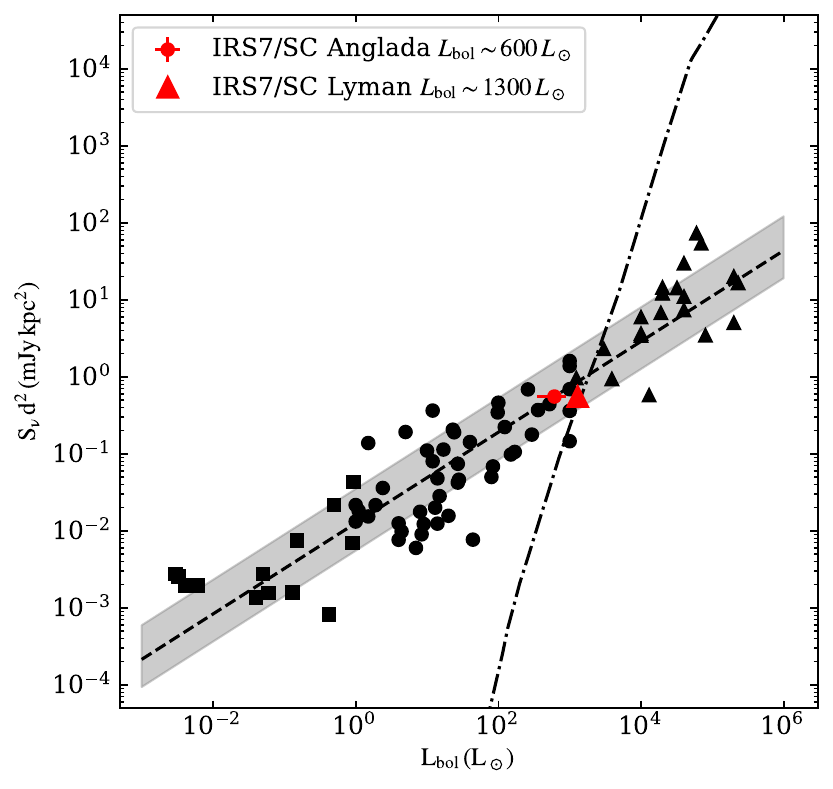}
        \caption{Correlation between bolometric luminosity and radio continuum luminosity at centimetre wavelengths (dashed line and shaded area). Squares represent very low luminosity ($L_\mathrm{bol}<1~L_\odot$), circles low luminosity ($1\leq L_\mathrm{bol}\leq1000~L_\odot$), and triangles high luminosity sources ($L_\mathrm{bol}>1000~L_\odot$). The dot-dashed line represents the expected radio luminosity of an optically thin region photo-ionised by Lyman continuum. The red circle corresponds to a $L_\mathrm{bol}\sim600~L_\odot$ assuming all radio flux comes from shocked-excited jet emission, whereas the red triangle corresponds to a $L_\mathrm{bol}\sim1300~L_\odot$ assuming all radio flux comes from an optically thin \ion{H}{II} region. Both have been calculated using $S_\nu d^2=0.55~\mathrm{mJy~kpc^2}$, with $\nu=8.19$~GHz. The figure is adapted from \citet{anglada2018}.}
        \label{fig:anglada_plot}
    \end{figure}

    To further investigate the nature of this source, we compared its bolometric and radio luminosities with those of known YSOs. Using Eq.~(28) from \citet{anglada2018} and our observations in the VLA X and C bands, i.e. $\nu=8.19$~GHz, $S_\nu=0.283$~mJy, and $d=1400$~pc (Table~\ref{tab:vla}), we estimated the bolometric luminosity assuming the entire radio emission arises from an ionised jet, yielding $L_\mathrm{bol}\sim600~L_\odot$ (red circle in Fig.~\ref{fig:anglada_plot}). Alternatively, if the centimetre radio flux originates from an optically thin \ion{H}{ii} region, then we obtain a rate of ionising photons (or Lyman photon rate) of $\log_{10}(\dot{N}_i/\mathrm{s^{-1}})\sim43.72$, assuming $T_e=10^4$~K \citep[see e.g.][]{kurtz1994,estalella1999}. The bolometric luminosity inferred from these ionising photons is then $L_\mathrm{bol}\sim1300~L_\odot$. Interpolating from Table~1 of \citet{thompson1984}, the corresponding Lyman photon rate matches that expected from a B2/B3 ZAMS star.

    The excellent agreement between the spectral classification inferred independently from the near-IR Br$\gamma$ line and from the radio continuum supports the interpretation of IRS~7/SC as a compact \ion{H}{ii} region powered by a young B2/B3 ZAMS star. Its UV radiation appears to be carving out and illuminating a surrounding PDR, as evidenced by the observed H$_2$ morphology and excitation conditions. Other sources in the vicinity of IRAS~18162-2048 can be ruled out as the main UV radiation field. For example, IRS~2 does not show any emission or absorption line in its near-IR $K$ band spectrum and no $3-6$~cm nor 3.3~mm detection. The main protostar IRAS~18162-2048 is deeply embedded in its parental core, with a thick accretion disc that may confine any possible UV radiation within the first few astronomical units from the protostar \citep{carrasco-gonzalez2012,anez2020}.
    All in all, IRS~7/SC seems to be the most likely source of UV radiation in the field and therefore represents an important transitional phase between a massive protostar and a fully developed \ion{H}{ii} region \citep[see e.g.][]{osorio1999}, providing a valuable benchmark for understanding the early radiative feedback in massive star formation (see also Sect.~\ref{sect:multigenerational}).

\subsection{Physical conditions of the PDR}\label{sect:cloudy}

    \begin{figure*}[!htb]
        \centering
        \includegraphics[width=0.28\textwidth]{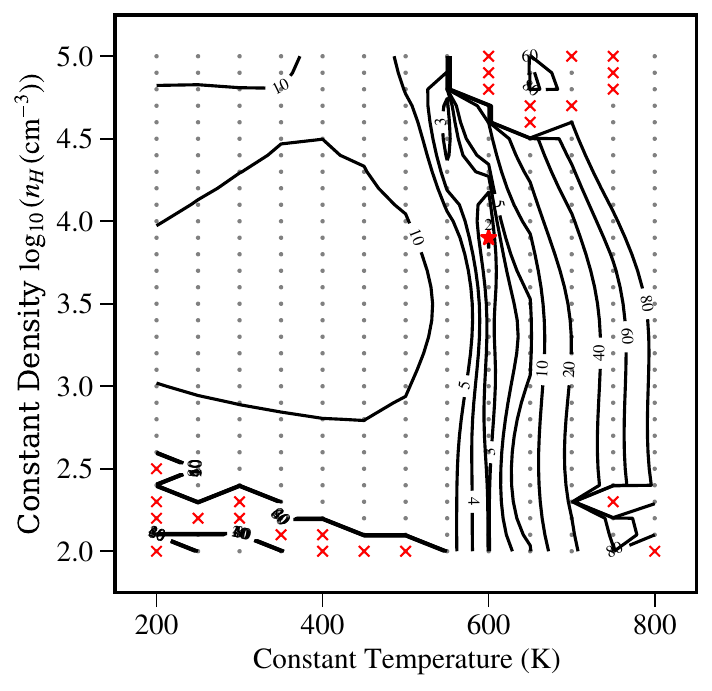}
        \includegraphics[width=0.63\textwidth]{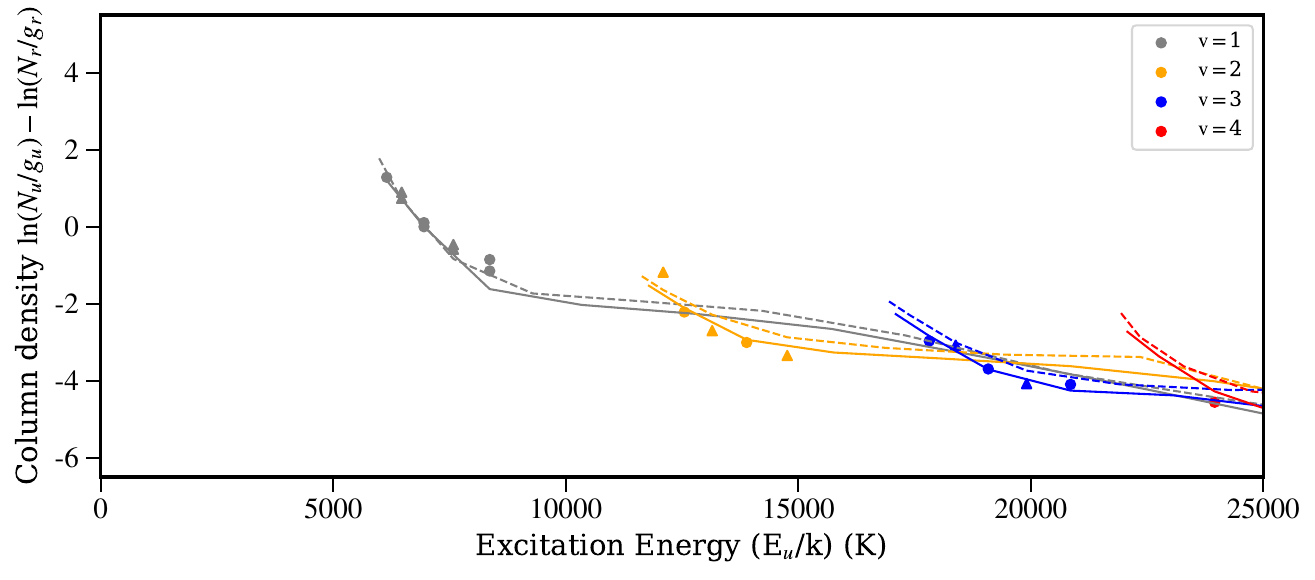}
        
        \caption{\textit{Left:} $\chi^2$ map for the grid run of constant density and constant temperature with Cloudy. The red star represents the model with the lowest $\chi^2$ of 1.94 (see text). The red crosses mark the grid values where the Cloudy code raised a failure flag of \texttt{failed assert} or \texttt{cloudy abort}. \textit{Right:} Ro-vibrational diagram of H$_2$ comparing our observations (markers) with the best Cloudy model (lines) for $T=600$~K and $\log_{10}(n_\mathrm{H}~/\mathrm{cm^{-3}})=3.9$. Both models and observations have been normalised to the column density over the statistical weight of the 1-0 S(1) line at $2.12~\mu $m (i.e. $N_r/g_r$). Markers and lines are colour-coded by vibrational level (v), and circles and solid lines represent ortho- transitions, whereas triangles and dashed lines are for para- transitions.}
        \label{fig:cloudy}
    \end{figure*}

    There is compelling evidence of the presence of a PDR in IRAS~18162–2048. Our observations indicate that IRS~7/SC is most likely a B2/B3 ZAMS star that has ignited its own compact \ion{H}{ii} region. The surrounding molecular gas shows clear signatures of UV-pumped H$_2$ emission, confirming the coexistence of ionised, atomic, and molecular phases typical of a PDR.

    To investigate the physical conditions governing this region, we used the spectral synthesis and photo-ionisation code Cloudy.\footnote{\url{https://gitlab.nublado.org/cloudy/cloudy/-/wikis/home}} Calculations were carried out with version 25.00 of Cloudy \citep[described by][]{gunasekera2025}, which includes significant improvements in its treatment of molecular processes and updated atomic and grain physics compared to the widely used versions 13 and 17 \citep{ferland2013,ferland2017}.

    Following a methodology similar to that of \citet{kaplan2017}, we explored the PDR’s physical properties using a grid of constant-temperature, constant-density models. Although this approach simplifies the geometry, it effectively reproduces the observed H$_2$ ro-vibrational level populations, suggesting that the emitting layer of the PDR can be characterised by relatively uniform conditions. We used the \texttt{grid} command in Cloudy to explore temperatures from 200 to 800~K in steps of 50~K and hydrogen densities from $\log_{10}(n_\mathrm{H}~/\mathrm{cm^{-3}}) = 2.0$ to $5.0$ in steps of 0.1 dex. The incident radiation field was modelled as a blackbody with an effective temperature of $T = 19~000$~K, consistent with a B3~V star, and a hydrogen-ionising photon rate of $\log_{10}(\dot{N}_i/\mathrm{s^{-1}})\sim43.72$ \citep{thompson1984}, as obtained from the radio observations. The projected separation between IRS~7/SC and the PDR, derived from our near-IR observations, is $4\arcsec$ (corresponding to 5600~au, or 0.027~pc), which we specified using the Cloudy command \texttt{radius 16.92} in units of $\log_{10}$ in centimetres. Standard interstellar medium abundances were adopted, including dust grains and polycyclic aromatic hydrocarbons (PAHs), by enabling the commands \texttt{grain Orion}, \texttt{grains PAH 3}, and \texttt{database H2}. To account for element depletion onto dust, we used the parametrisation of \citet{jenkins2009} with a depletion strength of $F^* = 0.5$. Cosmic background radiation and cosmic ray ionisation were included via the \texttt{CMB} and \texttt{cosmic ray background} options, respectively. The models were run with 3000 computational cells and with the \texttt{sphere} and \texttt{double} keywords to ensure proper convergence for optically thick geometries. A stopping criterion of $A_V = 20$~mag was imposed, motivated by the observed extinction ($A_V \sim 12$~mag; Sect.~\ref{sect:H2_PDR}). The calculations were set to iterate until convergence.

    A total of 403 models were generated. The goodness of the fit was quantified using $\chi^2 = \Sigma (\mathrm{\ln(F_\mathrm{obs}) - \ln(F_\mathrm{model})})^2\label{eq:chisq}$, where $F_\mathrm{obs}$ and $F_\mathrm{mod}$ are the observed and modelled line flux densities, respectively. Figure~\ref{fig:cloudy} (left panel) shows the resulting $\chi^2$ contour map. The best-fitting model ($\chi^2 = 1.94$) corresponds to $T = 600$~K and $\log_{10}(n_\mathrm{H}~/\mathrm{cm^{-3}}) = 3.9$, i.e. $n_\mathrm{H}=7.9\times10^3~\mathrm{cm^{-3}}$. The grid strongly constrains the gas temperature, while the density remains poorly constrained, a degeneracy also noted by \citet{kaplan2017} in their study of the Orion Bar. This behaviour arises because temperature affects the collisional excitation rates and the thermal population of the $v = 0$ ladder, which in turn determines how higher vibrational levels are populated via UV pumping. As a result, the models are highly sensitive to temperature variations, whereas density exerts a weaker influence. As mentioned in Sect.~\ref{sect:rv_diag}, the fitted $T_\mathrm{cold}$ component may be a better tracer of the kinetic temperature of the gas, and it is consistent with the temperature of the best Cloudy model.

    The right panel of Fig.~\ref{fig:cloudy} presents the ro-vibrational excitation diagram for the best-fit model compared with our de-reddened observations, both normalised to the $1-0$~S(1) line at $2.12~\mathrm{\mu m}$. The agreement is excellent, particularly in reproducing the observed `sawtooth' pattern characteristic of UV-pumped molecular hydrogen. Model-independent estimates of the radiation field strength can be expressed in units of the Habing field \citep{wolfire2022} as $G_0\sim0.5L_\mathrm{bol}/(4\pi l^2 1.6\times10^{-3})$, where $L_\mathrm{bol}$ is the stellar luminosity (in $\mathrm{erg~s^{-1}}$) and $l$ is the projected distance in cm between the ionising source and the PDR. Using $L \sim 1300~L_\odot \sim 5\times10^{36}~\mathrm{erg~s^{-1}}$ and $l \sim 5880~\mathrm{au} \sim 8.79\times10^{16}~\mathrm{cm}$, we estimated $G_0 \sim 1.6\times10^4$. For comparison, typical field strengths in benchmark PDRs such as the Orion Bar are $G_0 \sim 3\times10^4$ \citep{wolfire2022}, while \citet{berne2009} derived values in the range of $G_0 \sim 10^3-10^5$ for Monoceros~R2 from the combined analysis of PAH ionisation fractions and H$_2$ $0-0$~S(3)/S(2) line ratios. The derived parameters, $T = 600$~K, $n_\mathrm{H} \approx 8 \times 10^{3}~\mathrm{cm^{-3}}$, and $G_0\sim10^4$, are consistent with those found in other dense PDRs illuminated by early-type stars. The estimated radiation field in IRAS~18162–2048 is therefore consistent with a dense, strongly irradiated PDR powered by a B2/B3 ZAMS star.

    Our spectral coverage ($1.93-2.47~\mathrm{\mu m}$) only includes the near-IR ro-vibrational transitions of H$_2$ and thus misses the pure rotational $0-0$ transitions at mid-IR wavelengths ($3.3-28.2~\mathrm{\mu m}$). These lines, observable only with the \textit{James Webb} Space Telescope (JWST), would probe the colder gas component of the PDR and provide critical constraints on the gas density and heating balance. Therefore, follow-up JWST observations are warranted to fully characterise the physical structure and excitation conditions in the region IRAS~18162–2048.

\subsection{Multi-generational star formation in IRAS~18162-2048}\label{sect:multigenerational}

    The high-mass star-forming region IRAS~18162-2048 presents convincing evidence of multi-generational star formation, with young protostars, compact \ion{H}{ii} regions, and evolved feedback-driven structures coexisting within the same molecular environment. In addition to the well-studied collimated jet driven by the main protostar IRAS~18162-2048, several bubble-like cavities have now been identified across the region (see the H$_2$ continuum subtracted image shown in Fig.~16 from \citealt{bally2023} and Fig.~2 from \citealt{mohan2023obs}). These include bubble-like structures centred on IRS~4, IRS~3, and a third source southeast of MHO~2360 as well as the bubble revealed in this study around IRS~7/SC (see Fig.~\ref{fig:cont_H2_BrG}). Such a morphology is more consistent with the cumulative action of multiple photo-ionising sources over time than with purely shock-excited bow shocks from protostellar outflows. However, we cannot rule out that some of this molecular emission emerges from outflow shocked material.

    Our VLA detections provide further support for a more evolved stellar population embedded within the region. Both IRS~4 and IRS~3 are clearly detected at centimetre wavelengths, suggesting the presence of ionised gas. Although the lack of equivalent ALMA or near-IR integral field unit coverage prevents a full characterisation of these sources, the spatial coincidence between radio peaks and bubble H$_2$ structures strongly hints at past or ongoing \ion{H}{ii} regions. We also calculated the number of Lyman continuum photons for the two sources as done for IRS~7/SC (see Sect.~\ref{sect:discussion_irs7}). We obtained $\log_{10}(\dot{N}_i/\mathrm{s^{-1}})\sim43.82$ for IRS~4 and $\log_{10}(\dot{N}_i/\mathrm{s^{-1}})\sim43.83$ for the source in the proximity of IRS~3 (see Table~\ref{tab:vla}), both consistent with B2/B3 ZAMS stars. In fact, these sources may represent an older generation of stars, as they are clearly detected in the H$\alpha$ image of \citet[see their Fig.~1]{bally2023} and therefore had time to disperse their surroundings and lower the extinction. This structure should be considered in conjunction with the newly detected PDR around IRS~7/SC, which our combined near-IR and radio diagnostics favour to be a compact photo-ionised nebula. In this case, however, IRS~7/SC may be in a younger evolutionary stage in comparison to the aforementioned sources.

    Taken together, the presence of multiple ionised bubbles and compact radio sources indicates that several massive or intermediate-mass stars have already reached the stage of producing photo-ionising feedback. Meanwhile, the main protostar, IRAS~18162-2048, appears to be significantly younger. Estimates for massive protostars undergoing high accretion rates ($10^{-4}$–$10^{-3}~M_\odot~\mathrm{yr^{-1}}$) predict the development of a compact \ion{H}{ii} region at ages $\sim10^5$~yr \citep{osorio1999,cesaroni2005IAU,davies2011,mottram2011}. Although \citet{anez2020} suggested the presence of an incipient hyper-compact \ion{H}{ii} region towards the main protostar, observational evidence remains inconclusive. Our results are consistent with IRAS~18162-2048 still being younger than $\sim10^5$~yr and not in the ZAMS stage yet, i.e. prior to fully developing a stable photo-ionised region, while other sources in its vicinity have already reached this evolutionary stage.

    This age spread opens the possibility that a triggered star formation process operated in the region IRAS~18162-2048. The spatial distribution of the bubbles appears to surround the central region where the main protostar IRAS~18162-2048 resides. Although speculative at this stage, the geometry towards the main protostar IRAS~18161-2048 may be suggestive of compressed gas being funnelled inwards by previous generations of feedback-driven expansion. Such a sequence of older stars carving cavities followed by the collapse of swept-up material forming younger objects has been proposed in other high-mass star-forming regions \citep[e.g.][]{lee2007,thompson2012,liu2017,kim2025}. Nevertheless, the current data do not yet allow for a firm conclusion, and dedicated molecular-line kinematics studies and higher-resolution radio and IR spectro-imaging of the outer bubbles will be required to confirm or refute this scenario.

    In summary, the high-mass star-forming region IRAS~18162-2048 is emerging as a clustered sequential star-forming environment where feedback from earlier massive stars has shaped the cloud and may have influenced the birth of the next generation, including the main protostar IRAS~18162-2048. The system therefore provides an excellent laboratory to study the interplay between feedback and star formation in a massive protocluster context.

\section{Conclusions}\label{sect:conclusions} 

    We have presented near-IR integral field spectroscopic observations of the high-mass star-forming region IRAS~18162–2048 obtained with VLT/SINFONI and complemented with new VLA X and C band data and ALMA band~3 observations. Our study resolved the inner $\sim$10000~au of the system, revealing a complex interplay between massive and intermediate-mass YSOs, ionised and molecular gas, and UV-irradiated material. Through spectral and morphological diagnostics, we identified clear evidence of a compact \ion{H}{ii} region and a PDR excited by a B2/B3 type ZAMS star, most likely by IRS~7/SC. We summarise our main conclusions in the following:

\begin{itemize}
    \item The SINFONI $K$ band continuum map reveals two main IR sources, IRS~2 and IRS~7, consistent with previous detections by \citet{yamashita1987,tamura1991,stecklum1997}. The primary source, IRAS~18162-2048, remains undetected up to 2.47~$\mathrm{\mu m}$, indicating very high extinction towards the main protostar (Fig.~\ref{fig:cont_H2_BrG}).

    \item The IR source IRS~7 shows a peculiar Br$\gamma$ profile with a narrow emission core superimposed on a broad absorption component, and this is consistent with a B2–B3 ZAMS star following the classification scheme from \citet{bik2005b}. The lack of \ion{He}{i} and \ion{C}{iv} lines rules out O spectral types. We interpret the emission component of the Br$\gamma$ line as evidence of a nascent compact \ion{H}{ii} region surrounding IRS~7 and the absorption component as its photosphere (Fig.~\ref{fig:BrG_zoom}).

    \item Strong H$_2$ ro-vibrational emission was detected, exhibiting the characteristic `sawtooth' pattern in the excitation diagram, which is clear evidence of UV radiation typical of PDRs, rather than the shock excitation. The $1-0$~S(1)/$2-1$~S(1) ratio ($\leq$5.9) further supports this interpretation and is consistent with dense UV-irradiated molecular gas. Pixel-by-pixel H$_2$ $1-0$~Q(3)/$1-0$~S(1) ratios yield an average visual extinction of $A_V\sim12$~mag across the main bow-shaped H$_2$ structure, consistent with previous estimates. This moderate extinction enables efficient UV penetration, sustaining the observed H$_2$ emission. The radial velocity map reveals bulk velocities of $\sim1\,\mathrm{km\,s^{-1}}$, further evidencing the dominance of UV radiation. See Figs. \ref{fig:extinction_excitation} and \ref{fig:rv_diag}.

    \item New high-resolution VLA X and C band observations detected the compact source coincident with IRS~7 (also known in the radio as `stationary condensation'), with a flux density of $0.283\pm0.015$~mJy (Fig.~\ref{fig:vla_cx}). For the first time, IRS~7/SC is detected in millimetre wavelengths in our ALMA band~3 observations, with a flux density of $0.457\pm0.310$~mJy (Fig.~\ref{fig:alma_band3}). The spectral index between $3-6$~cm and 3.3~mm ($\alpha=0.19\pm0.27$) is consistent with free-free emission. The derived Lyman continuum photon rate for the centimetre regime (Fig.~\ref{fig:anglada_plot}) corresponds to a B2/B3 ZAMS star, in excellent agreement with the near-IR spectral diagnostics.

    \item Using the 25.00 Cloudy code \citep{gunasekera2025}, we modelled the H$_2$ ro-vibrational populations assuming a constant density and temperature. The best-fitted model yielded $T_\mathrm{gas}=600$~K and $n_\mathrm{H}=7.9\times10^3~\mathrm{cm^{-3}}$, reproducing the observed `sawtooth' excitation pattern (Fig.~\ref{fig:cloudy}). A model-independent estimate of the UV radiation field gave $G_0\sim1.6\times10^4$ in Habing units, which is similar to estimates inferred for benchmark PDRs such as the Orion Bar and Monoceros~R2. This supports the interpretation of IRS~7 as a young massive star photo-ionising and photo-dissociating its surroundings.

    \item Large and small scales near-IR and radio observations revealed multiple compact bubbles associated with sources such as IRS~7, IRS~4, IRS~3, and the structure southeast to MHO~2360, indicating previous episodes of massive-star feedback and the presence of \ion{H}{ii} regions. The morphology and spatial distribution of these bubbles suggest that earlier generations of stars influenced the cloud structure and potentially helped trigger subsequent star-formation events. While further high-resolution observations are required to firmly establish this scenario, IRAS~18162–2048 stands out as an active, multi-generational, feedback-rich star-forming complex.
\end{itemize}

\begin{acknowledgements}
    The authors acknowledge fruitful discussions with K.F. Kaplan, M. Pound, and M. Wolfire. The authors thank the anonymous referee for their constructive report. R.F., G.A., J.F.G, M.O., G.B.C., F.P.M, G.A.F acknowledge support from the grants PID2023-146295NB-I00, and from the Severo Ochoa grant CEX2021-001131-S funded by MCIN/AEI/ 10.13039/501100011033 and by ``European Union NextGenerationEU/PRTR''. G.B-C also acknowledges support from grant PRE2018-086111, funded by MCIN/AEI/ 10.13039/501100011033 and by `ESF Investing in your future'. A.F.P.M acknowledges support from grant PRE2021-100926, funded by MCIN/AEI /10.13039/501100011033 and by `ESF Investing in your future'. A.C.G. acknowledges support from PRIN-MUR 2022 20228JPA3A “The path to star and planet formation in the JWST era (PATH)” funded by NextGeneration EU and by INAF-GoG 2022 “NIR-dark Accretion Outbursts in Massive Young stellar objects (NAOMY)” and Large Gran INAF-2024 “Spectral Key features of Young stellar objects: Wind-Accretion LinKs Explored in the infraRed (SKYWALKER). JMMS acknowledges financial support from the Spanish Ministry of Science and Innovation under grant PID2022-136828NB-C41/AEI/10.13039/501100011033/ERDF/EU and through the Mar\'ia de Maeztu award to the ICCUB (CEX2024-001451-M). R.G.M acknowledges support from DGAPA-UNAM-PAPIIT project IN105225. C.C.-G. acknowledges support from UNAM DGAPA-PAPIIT grant IG101224. G.A.F also gratefully acknowledges the Deutsche Forschungsgemeinschaft  (DFG) for funding through SFB 1601 ``Habitats of massive stars across cosmic time'' (sub-project B1) and from the University of Cologne and its Global Faculty programme.
    
    This paper makes use of the following ALMA data: ADS/JAO.ALMA\#2012.1.00441.S. ALMA is a partnership of ESO (representing its member states), NSF (USA) and NINS (Japan), together with NRC (Canada), NSC and ASIAA (Taiwan), and KASI (Republic of Korea), in cooperation with the Republic of Chile. The Joint ALMA Observatory is operated by ESO, AUI/NRAO and NAOJ. This study is also based on observations made under project 19A-321 with the VLA of NRAO. The National Radio Astronomy Observatory is a facility of the National Science Foundation operated under cooperative agreement by Associated Universities, Inc.
\end{acknowledgements}

%
%

\bibliographystyle{aa}
\bibliography{phd_bibliography.bib}

\end{document}